\documentclass[review]{elsarticle}

\usepackage{lineno,hyperref}
\usepackage{setspace}
\modulolinenumbers[5]

\journal{Journal of \LaTeX\ Templates}

\usepackage{color}
\usepackage{xcolor}
\usepackage{comment}
\usepackage{amsmath,amssymb}

\def\int{\textit{INTEGRAL}}

\begin{document}

\begin{frontmatter}

\title{INTEGRAL view on Cataclysmic Variables and Symbiotic
Binaries}

\author{Alexander Lutovinov\corref{cor1}}%
\ead{aal@iki.rssi.ru}

\address{Space Research Institute, Profsoyuznaya str. 84/32, 117997 Moscow, Russia\\
Moscow Institute of Physics and Technology, Dolgoprudny, Moscow Region, 141700, Russia}

\author{Valery Suleimanov}
\address{Institute for Astronomy and Astrophysics, University of
T\"ubingen, Sand 1, 72026 T\"ubingen, Germany\\
Space Research Institute, Profsoyuznaya str. 84/32, 117997 Moscow, Russia
\\ Kazan Federal University, Kremlevskaya 18, 420008 Kazan, Russia}

\author{Gerardo Juan Manuel Luna}
\address{CONICET-Universidad de Buenos Aires, Instituto de Astronomía y
Física del Espacio (IAFE), Av. Inte. G\"uiraldes 2620, C1428ZAA, Buenos
Aires, Argentina \\
Universidad de Buenos Aires, Facultad de Ciencias Exactas y Naturales,
Buenos Aires, Argentina \\
Universidad Nacional de Hurlingham, Av. Gdor. Vergara 2222, Villa Tesei, Buenos Aires, Argentina. }

\author{Sergey Sazonov}
\address{Space Research Institute, Profsoyuznaya str. 84/32, 117997
Moscow, Russia \\
Moscow Institute of Physics and Technology, Dolgoprudny, Moscow Region, 141700, Russia}

\author{Domitilla de Martino}
\address{INAF - Capodimonte Astronomical Observatory Naples
Via Moiariello 16, I-80131 Naples, Italy}

\author{Lorenzo Ducci}
\address{Institute for Astronomy and Astrophysics, University of
T\"ubingen, Sand 1, 72026 T\"ubingen, Germany \\ ISDC Data Center for
Astrophysics, Universit\'e de Gen\`eve, 16 chemin d'\'Ecogia, 1290
Versoix, Switzerland}

\author{Victor Doroshenko}
\address{Institute for Astronomy and Astrophysics, University of
T\"ubingen, Sand 1, 72026 T\"ubingen, Germany \\ Space Research Institute,
Profsoyuznaya str. 84/32, 117997 Moscow, Russia}

\author{Maurizio Falanga}
\address{International Space Science Institute, Hallerstrasse 6, CH-3012
Bern, Switzerland}





\begin{abstract}
Accreting white dwarfs (WDs) constitute a significant fraction of the hard
X-ray sources detected by the \int\ observatory. Most of them are magnetic
Cataclysmic Variables (CVs) of the intermediate polar (IP) and polar types,
but the contribution of the Nova-likes systems and the systems with
optically thin boundary layers, Dwarf Novae (DNs) and Symbiotic Binaries (or
Symbiotic Stars, SySs) in quiescence is also not negligible. Here we present
a short review of the results obtained from the observations of cataclysmic
variables and symbiotic binaries by \int. The highlight results include the
significant increase of the known IP population, determination of the WD mass
for a significant fraction of IPs, the establishment of the luminosity
function of magnetic CVs, and uncovering origin of the Galactic ridge X-ray
emission which appears to largely be associated with hard emission from
magnetic CVs.
\end{abstract}


\end{frontmatter}

\nolinenumbers

\section{Introduction}

Cataclysmic variables (CVs) is a broad class of close binary systems with an
accreting white dwarf (WD) as a primary (see a detailed review in the book by
B.\,Warner \cite{Warner03}). The virial temperature of protons at the WD
surface is about a few tens of keV, so the optically thin accretion flow
shocked above the WD surface, inevitably produces an optically thin free-free
radiation with the comparable temperatures. Therefore, a significant fraction
of CVs appear as hard X-ray sources.

The first model describing this process in the case of the spherical
accretion on the WD surface was proposed by \cite{Aizu73}. Soon it was
recognized that this model is oversimplified and far from the reality. The
local shock areas near magnetic poles, which are shaped by the strong
magnetic field of the WD, were proposed as the most likely origin of the
observed hard X-ray radiation in CVs (see, e.g. \citep{FPR76}.) Almost at the
same time it was found that the well known CV AM\,Her was detected with the
{\it Uhuru} observatory as an X-ray source (3U\,1809+50)
\cite{1974ApJS...27...37G}. The source also showed a significant optical
polarization \cite{Tapia77}, indicating the presence of a strongly magnetized
WD in the system, and thus supported the hypothesis that the observed hard
X-ray emission was originating in the hot plasma heated by the shock above
the WD surface. For this particular object (AM\,Her) the spin period is equal
to the orbital period, therefore all similar CVs are named synchronous polars
(from the polarized optical emission) or AM\,Her type stars. There is no
accretion disc in such systems and the magnetic field strength of the WD
($10^7-10^8$\,G) is sufficient to synchronize the spin period with the
orbital one. Note that the synchronization is possible only for relatively
compact systems, so most polars have relatively short orbital periods $<5$\,h
(see, e.g., \cite{Norton.etal:04}).

The first objects belonging to another type of magnetic CVs, the intermediate
polars (IPs), were discovered as X-ray sources in the late of 70th (TV\,Col
\cite{Charles.etal:79} and AO\,Psc \cite{Griffiths.etal:80}). Spin periods of
WDs in such systems are much shorter compared to orbital ones, and most of
them did not display any optical polarization (see, e.g.,
\cite{2015SSRv..191..111F}). Unlike polars, in IPs systems there was clear
evidence for presence of an accretion disc (see details in \cite{Warner03}).
The magnetic field strength in these systems is believed to be lower than in
polars ($<1-3\times10^6$\,G), thus allowing the formation of an accretion
disc. It is also not strong enough to synchronize the WD rotation with the
orbital period. As a consequence, the accretion disk is truncated at some
radius defined as the magnetospheric radius, $R_{\rm m}$, where the magnetic
and accretion pressures balance. Surprisingly, this subclass of the magnetic
CVs has been found to be the most numerous among the hard X-ray emitting CVs
discovered by \int.

A crucial parameter of magnetic CVs, which can be learned from hard X-ray
observations, is a mass of the WD. Indeed, the temperature below the shock is
defined by a compactness of the WD, i.e., by the ratio of its mass and
radius. On the other hand, the relation between mass and radius of the WD is
well established (see, e.g. \cite{N72}). Therefore, it is possible to
estimate the WD mass based on the observed plasma temperature, which in turn
can be derived from the approximation of its hard X-ray spectrum with an
optically thin emission as the bremsstrahlung model (see, e.g.,
\cite{Rothschild.etal:81} for the first results).  Later more detailed models
of the so-called post-shock region (PSR) where the observed emission is
produced were developed (\cite{Wu.etal:94, WB96, Cropper.etal:99,
Canalle.etal:05, Saxton.etal:07}) and used to estimate the WD masses in
magnetic CVs  (see, e.g., \cite{CRW98, Ramsay:00}).

Note, that these early works used a relatively soft ($E < 20$\,keV) part of
IP X-ray spectra measured with the {\it Ginga} observatory and {\it RXTE}/PCA
instrument. Later, the same approach was applied to observations in hard
X-rays with the {\it RXTE}/HEXTE, {\it Swift}/BAT  instruments and with the
{\it Suzaku} observatory \citep[see, e.g., ][]{SRR05,Brunschweiger.etal:09,
Yuasa.etal:10}. A number of articles was dedicated to measure the masses of
individual IPs or smaller samples of objects (see, e.g.,
\cite{Anzolin.etal:09, Bernandini.etal:12, Bernandini.etal:13,
Bernandini.etal:15, Bernandini.etal:18, Bernardini.etal:19}). Recently, the
high sensitivity of the {\it NuSTAR} observatory in hard X-rays has been
exploited for the same purposes (see, e.g., \cite{Shaw.etal:18, Wada.etal:18,
SDW19}). The {\it INTEGRAL} observatory (\cite{Winkler.etal:03}) played also
an important role in the investigations of CVs. In particular, hard X-ray
spectra in many IP systems obtained by the IBIS telescope were used to
determine WD masses, starting from the early papers
\cite{Revnivtsev.etal:04b, Falanga.etal:05}.

Obviously, the determination of parameters of individual IPs is possible only
for relatively close, bright sources. At the same time an extended hard X-ray
emission, registered from the Galactic Center and Galactic ridge, can be
explained as collective emission of a large number of unresolved CVs. This
hypothesis, initially proposed by \cite{Revnivtsev.etal:06}, was later
confirmed in number of  papers \cite{Sazonov.etal:06, Revnivtsev.etal:08,
Hailey.etal:16}. Thus, the study of the Galactic ridge emission is tightly
connected with the analysis of individual CVs properties, and the {\it
INTEGRAL} observatory contributed significantly in this area.

It is important to note that the magnetic CVs are not the only WDs capable of
producing hard X-ray emission (see e.g. \cite{Mukai:17}). For instance, hard
X-ray emission from Dwarf Novae (DN) was already discovered at the end of 70s
(see, e.g.\cite{Swank.etal:78, Heise.etal:78}) during their quiescent states.
These results were interpreted as radiation of the optically thin boundary
layer between the WD surface and the accretion disc feeding the WD at low
mass accretion rates \cite{PS79, Tylenda81}. The system SS\,Cyg is the best
characterized DN in hard X-rays at the present time (see, e.g.
\cite{McGowan.etal:04, BR12}). In addition to DNs, Novalike systems have been
detected in the hard X-rays although some of them  are disputed to be
magnetic (e.g. IGR J12123-5802 \cite{Bernandini.etal:13}; TW Pic
\cite{Norton.etal:00}) .

Symbiotic Stars (SySs) are a quite small subclass of binary systems, which
emit in X-rays due to the accretion onto compact objects. The donor
star in such systems is a red giant, and
it gives a main contribution in the optical luminosity. A compact
object (typically a WD, but there are several systems with
neutron stars) accretes the matter from donor wind producing
ultraviolet and X-ray emission extending down to the blue optical
spectral band (see recent review \cite{Munari:19} and references therein).
Below we will focus on the systems with WDs as an accretor. The
SySs with neutron stars are characterised in a separate review paper
in this volume.

The first SySs with accreting WDs were discovered by the {\it Einstein}
observatory at the beginning of the 80s \cite{Anderson.etal:81,
Cordova.etal:81, Allen:81}. The origin of the X-ray radiation in such systems
can be similar to that in CVs although they show also some remarkable
difference with respect to them. In particular, there are super-soft sources
with permanent thermonuclear burning on the WD surface, like the AG\,Dra
system. Boundary layers between accretion disks and WDs are considered as
most probable sources of X-ray radiation in less luminous systems. The hard
X-ray emission of RT Cru discovered by \int\ (\cite{Chernyakova.etal:05}) was
proposed to originate from an optically thin boundary layer.

Here we present a detailed review of the results concerning various types of
CVs, obtained with the \int\ observatory. The review includes theoretical
explanations discussed in connection with properties of individual sources
and their populations such as the luminosity function and contribution to the
Galactic ridge emission.

\section{CVs and SySs observed by \int}

A significant part of the {\it INTEGRAL} observational program is dedicated
to surveys of different regions of the Galactic plane and its regular scans
with the purpose to map the Galaxy, detect new sources and to study in
details all detected objects. The main instrument of the observatory for such
investigations is the ISGRI detector of the IBIS telescope, ISGRI/IBIS (the
Imager on Board the INTEGRAL Satellite / INTEGRAL Soft Gamma Ray Imager,
\cite{Lebrun.etal:03}). During more than 15 years {\it INTEGRAL} performed
the deepest hard X-ray survey of the Galactic plane, which resulted in the
discovery and characterization of many new and previously known sources which
have been reported in many publications and particularly, the IBIS/ISGRI
catalogues (see, e.g., \cite{Krivonos.etal:12, Bird.etal:16,
Krivonos.etal:17}).

These surveys performed with \int, besides producing general catalogues, have
allowed to study different classes of sources, including the CVs. The first
review of CVs, observed with {\it INTEGRAL}, was published by
\cite{Barlow.etal:06} and contained 19 known and newly discovered CVs. Most
of them were IPs, but a couple of AM Her type systems and one dwarf Nova (SS
Cyg) were also included. In a subsequent study, 22 objects were reported
\cite{Landi.etal:09} and most of them were again IPs. In both works the hard
X-ray spectra of the sources were fitted with bremsstrahlung and power-law
models. For 11 sources the soft \emph{Swift}/XRT spectra were also analysed
in the follow-up work by  \cite{Gehrels.etal:04}.

The Galactic plane surveys were extended further to the all-Sky surveys (see,
e.g., \cite{Krivonos.etal:12, Krivonos.etal:17}), which allowed to expand
significantly the list of known hard X-ray sources. The current version of
the general {\it INTEGRAL} source
catalogue\footnote{https://www.isdc.unige.ch/integral/catalog/latest/catalog.html}
contains a few tens of objects, identified as various types of CVs and
Symbiotic stars (see Table\,\ref{tab:int_compil}). We added some objects only
recently identified as CVs, and marked as LMXB a source, which was
erroneously identified earlier as CV \cite{deMartino.etal:10}.

\begin{table*}
\caption{List of CVs and SySs observed by {\it INTEGRAL}.
 \label{tab:int_compil} }
{\footnotesize
\begin{center}
\begin{tabular}[c]{ c | l l c c  l l }
\hline
N & Source ID & 	Name   & $\alpha_{2000}$ & $\delta_{2000}$ & Type & Ref  \\ [1mm]
\hline
\hline
1 & J002324.0+614132 &	V1033 Cas$^*$	 &	00:22:57.6 & +61:41:08 & IP & \cite{Bikmaev.etal:06}\\
2 & J002848.9+591722 & V709 Cas	 & 	00:28:48.9	& +59:17:22 & IP &\\
3 & J005524.0+461211 &	V515 And     &  00:55:24.0	& +46:12:11 & IP & \cite{Bikmaev.etal:06}\\
4 & J025604.1+192624 &	XY Ari	&	02:56:04.1	& +19:26:24 & IP & \\
5 & J033111.8+435417 &	GK Per	&  03:31:11.8 &	+43:54:17 & IP, DN, Nova &\\
6 & J045707.4+452751	& IGR J04571+4527$^*$ &	04:57:07.0 &	+45:27:48 & IP & \cite{TH13}\\
7 & J050227.5+244523	& V1062 Tau & 05:02:27.5 &	+24:45:23 & IP &\\
8 & J051029.3-691012 &	IGR J05104-6910$^*$ & 05:10:29.3 & -69:10:12 & DDP? & \cite{Haberl.etal:17}  \\
9 & J052522.5+241332 &	RX J0525.3+2413	 & 05:25:22.7 &	+24:13:33 & IP & \cite{Bernandini.etal:15} \\
10 & J052924.0-324840 &	TV Col & 05:29:24.0 &	-32:48:40 & IP &\\
11 & J053450.5-580139 &	TW Pic	   & 05:34:50.5	& -58:01:39 &	IP? & \\
12 & J054248.9+605132 &	BY Cam	& 05:42:48.9 &	+60:51:32 & AM & \\
13 & J055807.4+535443 &	V405 Aur & 05:58:07.4 &	+53:54:43 & IP &\\
14 & J061033.6-484426 &	V347 Pup &	06:10:33.6 & -48:44:26 & NL & \cite{Buckley.etal:90}\\
15 & J062516.1+733437 & MU Cam  & 06:25:22.0 &	+73:36:07	& IP &\\
16 & J073237.5-133104 & V667 Pup &	07:32:37.5 & -13:31:04 & IP &\\
17 & J074623.3-161341 & SWIFT J0746.3-1608 & 07:46:23.3 &	-16:13:41 & IP & \cite{Bernardini.etal:19} \\
18 & J075117.2+144421 & PQ Gem &  07:51:17.4 &	+14:44:25 & IP &\\
19 & J080108.2-462244 &	1RXS J080114.6-46232 & 08:01:17.0 & -46:23:27 & IP & \cite{Bernandini.etal:17} \\
20 & J083848.9-483125 &	IGR J08390-4833$^*$  & 08:38:49.1 &	-48:31:25 & IP & \cite{Revnivtsev.etal:09} \\
21 & J095750.7-420838 & SWIFT J0958.0-4208 & 09:57:50.6 &	-42:08:36 & IP & \cite{Masetti.etal:13, Bernandini.etal:17} \\
22 & J101050.4-574648 & IGR J10109-5746$^*$ & 10:11:03.0	& -57:48:15 & SyS &\cite{Masetti.etal:06b}\\
23 & J114338.2+714121 &	DO Dra & 11:43:38.3	& +71:41:20 & IP &\\
24 & J121226.0-580023	& IGR J12123-5802 &	12:12:26.2 & -58:00:21 & NL or IP? & \cite{Bernandini.etal:13} \\
25 & J123456.0-643400 &	RT Cru & 12:34:54.7	& -64:33:56 & SyS &\\
26 & J123816.3-384246 &	V1025 Cen & 12:38:16.3 &	-38:42:46 & IP& \\
 27  & J124853.5-624305 & IGR J12489-6243	& 12:48:53.5 &	-62:43:05 & CV? &
   \cite{Tomsick.etal:12,Fortin.etal:18} \\	
28 & J125224.4-291457 & EX Hya & 12:52:24.4 & 	-29:14:57 & IP &\\
29 & J140907.5-451717 &	V834 Cen	& 14:09:07.5 &	-45:17:17 & AM & \\
30 & J140846.0-610754 &	IGR J14091-6108$^*$ &	14:08:46.0	& -61:07:54 & IP & \cite{Tomsick.etal:16} \\
31 & J142507.7-611858 &	IGR J14257-6117$^*$ &	14:25:07.6	& -61:18:58 & IP &
\cite{Masetti.etal:13,Bernandini.etal:18} \\
32 & J145338.0-552223 &	IGR J14536-5522$^*$ &  14:53:41.1	& -55:21:39 & AM & \cite{Masetti.etal:06} \\
33 & J150918.8-665100 &	IGR J15094-6649$^*$ &  15:09:26.0	& -66:49:23 & IP & \cite{Masetti.etal:06} \\
34   & J152915.8-560947 &	IGR J15293-5609	& 15:29:15.8 & -56:09:47 & CV? &
   \cite{Tomsick.etal:12, Fortin.etal:18} \\	
35 & J154814.7-452840	& NY Lup	& 15:48:14.7 &	-45:28:40 & IP & \\
36   & J155246.9-502953 &	IGR J15529-5029	&	15:52:46.9	& -50:29:53	& CV? & \cite{Tomsick.etal:09} \\
37 & J155930.2+255514 &	T CrB &	  15:59:30.2	& +25:55:13 & SyS  &\cite{Masetti.etal:08}\\
38 & J161642.0-495700 & IGR J16167-4957$^*$  & 16:16:37.7 &	-49:58:44 & IP & \cite{Masetti.etal:06} \\
39 & J163553.8-472541 &	IGR J16358-4726  & 16:35:53.8 &	-47:25:41 & SyS & \cite{Nespoli.etal:08}\\
40 & J165001.2-330658 &	IGR J16500-3307$^*$ & 16:49:55.6	& -33:07:02 & IP & \cite{Masetti.etal:08} \\
41 & J165443.7-191630 &	IGR J16547-1916$^*$ &	16:54:43.7 & 	-19:16:31 & IP & \cite{Masetti.etal:10} \\
42 & J170120.8-430531 & IGR J17014-4306 & 	17:01:28.2 & -43:06:12 & IP &
\cite{Masetti.etal:13,Bernandini.etal:17} \\
43 & J171236.5-241445 &	V2400 Oph	&  	17:12:36.5 &	-24:14:45 & IP &\\
44 & J171930.0-410100 &	IGR J17195-4100$^*$ &	17:19:35.9 &	-41:00:54 & IP & \cite{Butters.etal:08} \\
45 & J172000.0-311600 & IGR J17200-3116 & 17:20:05.9 & -31:17:00 & SyS? &  \cite{Fortin.etal:18} \\
46 & J173024.0-060000 &	V2731 Oph$^*$ &	17:30:21.5	& -05:59:34 & IP & \cite{Gaensicke.etal:05} \\
47 & J173159.8-191356 &	V2487 Oph  & 17:31:59.8 &	-19:13:56 & Nova, IP? & \cite{Hernanz14} \\
48 & J174017.7-290356 & AX J1740.2-2903 & 	17:40:16.1 &	-29:03:38 & IP & \cite{HG10} \\
49   & J174026.9-365537 & IGR J17404-3655	&	17:40:26.9	& -36:55:37	& CV or IP? &\cite{Clavel.etal:19, Fortin.etal:18} \\
\hline
\end{tabular}
\end{center}
}
\end{table*}

\pagebreak
{\footnotesize Table 1 (continue)}
{\footnotesize
\begin{center}
\begin{tabular}[c]{ c | l l c c l l }
\hline
N & Source ID & 	Name   & $\alpha_{2000}$ & $\delta_{2000}$ & Type & Ref  \\[1mm]
\hline
\hline
50 & J174624.0-213300 &	1RXS J174607.8-21333 &  17:46:03.2 &	-21:33:27 & SyS & \cite{Masetti.etal:08}\\
51 & J174955.4-291920 &	CXOGBS J174954.5-294335 & 17:49:55.4 &	-29:19:20 & IP & \cite{Johnson.etal:17} \\
52 & J175834.6-212322 & IGR J17586-2129$^*$	& 17:58:34.6 & -21:23:22 & SyS? & \cite{Fortin.etal:18} \\
53 & J180035.7+081106 &	V2301 Oph	& 18:00:35.8 &	+08:11:06 & AM & \\
54 & J180450.6-145450 & IGR J18048-1455 &	18:04:39.0	& -14:56:47 & IP & \cite{2012AstL...38..629K, Middleton.etal:12} \\
55 & J180900.9-274214	& IGR J18088-2741 & 18:09:01.0	& -27:42:14	& IP? &\cite{Tomsick.etal:16b, Rahoui.etal:17} \\
56   & J181504.0-105132 &	IGR J18151-1052	& 	18:15:03.8 &	+10:51:35 & IP?  & \cite{2012AstL...38....1L,Masetti.etal:13}\\
57 & J181613.3+495204 &	AM Her	& 	18:16:13.3 &	+49:52:04 & AM  &\\
58 & J181722.2-250842 &	IGR J18173-2509$^*$ &   18:17:22.3 &	-25:08:43 & IP & \cite{Nichelli.etal:09} \\
59 & J181826.4-235248 &	IGR J18184-2352$^*$ & 18:18:26.4 &	-23:52:48 & CV? & \cite{Krivonos.etal:17} \\
60 & J182920.2-121251 &	IGR J18293-1213$^*$ &	18:29:20.2 &	-12:12:51 & IP? & \cite{Clavel.etal:16} \\
61 & J183049.9-123219 &	IGR J18308-1232$^*$ & 18:30:49.9 &	-12:32:19 & IP & \cite{Masetti.etal:09} \\
62   & J1832.3-0840 & AX J1832.3-0840 &	 18:32:19.3	& -08:40:30 & IP & \cite{Masetti.etal:13}\\
63 & J185502.2-310948 & V1223 Sgr &	 18:55:02.2	& -31:09:48 & IP &\\
64 & J190713.6-204554 & V1082 Sgr$^*$	& 19:07:13.7 &	-20:45:54	& preCV & \cite{Tovmassian.etal:18} \\
65   & J192445.6+501414 &	CH Cyg	&	19:24:45.6	& +50:14:14	& SyS & \\
66 & J192627.0+132205 & IGR J19267+1325$^*$ &  19:26:27.0	 & +13:22:05 & IP & \cite{Evans.etal:08} \\
67 & J194011.5-102525 & V1432 Aql & 19:40:11.5 &	-10:25:25 & AM &\\
68 & J195511.0+004442 & IGR J19552+0044 &  19:55:12.5 &	+00:45:37 & pre-polar or IP  & \cite{Bernandini.etal:13,Tovmassian.etal:17} \\
69 & J195814.9+323018 &	V2306 Cyg & 19:58:14.9 &	+32:30:18 & IP &\\
 70  & J201531.4+371116 &	RX J2015.6+3711	&	20:15:31.4 &	+37:11:17 & CV + blazar&
   \cite{Bassani.etal:14}	\\
 71  & J210933.8+432046 &	IGR J21095+4322	& 21:09:24.2 &	+43:19:36 & CV? & \cite{Halpern.etal:18}\\
72 & J211352.8+542215 &	1RXS J211336.1+54222 &  21:13:35.4 & +54:22:33 & IP & \cite{Bernandini.etal:17} \\
73 & J212344.8+421802 &	V2069 Cyg & 21:23:44.8 &	+42:18:02 & IP &\\
74 & J213330.0+510531 & IGR J21335+5105$^*$ & 21:33:30.0 &	+51:05:31 & IP &\\
75 & J214242.7+433509 &	SS Cyg	 &  21:42:42.8 &	+43:35:10 & DN &\\
76 & J221755.4-082105 &	FO Aqr	&  22:17:55.4 &	-08:21:05 & IP &\\
77 & J225518.0-030943 &	AO Psc	 &  22:55:18.0	& -03:09:43 & IP &\\
78 & J232954.3+062811  &   EI Psc &	 23:29:54.3	& +06:28:11 & DN &\\
 \hline
\end{tabular}
\end{center}
Note: $^*$ - discovered by {\it INTEGRAL}, IP - an intermediate polar, AM - a
polar, type AM Her, DN - a dwarf Nova, NL - a Nova-like star, preCV - a
precataclysmic CV, SyS - a symbiotic star, DDP - a double degenerate polar, a
secondary star is also WD, LMXB - a low mass X-ray binary.
}

\bigskip

In total, the table includes 78 objects: most of them are IPs and
candidates (51 sources), seven sources are polars or AM\,Her type systems,
six symbiotic stars, two dwarf Novae (SS\,Cyg and EI\,Psc), one
pre-cataclysmic variable star (V1082\,Sgr), one nova-like star
(V347\,Pup), and one possible double degenerate polar (IGR\,J05104-6910).
It is possible that the last system is a close binary consisting of two
WDs. There are also six candidate CVs which possibly are magnetic systems
due to their hard X-ray radiation, and two candidate symbiotic stars.
All references associated with the classification or re-classification of
individual objects are presented in Table\,\ref{tab:int_compil}.

It is important to emphasize that a large fraction of CVs in the Table,
namely 21, was either discovered or identified as such with the help of
{\it INTEGRAL}. These sources are marked with asterisks. Note that some of
them had already standard names for variable stars but were only identified as
CVs by X-ray observations.

The identification of all hard X-ray sources started immediately
after their discoveries  by {\it INTEGRAL}, and it is obvious that new CVs
were also a part of such programs (see, e.g., \cite{Masetti.etal:06,
Bikmaev.etal:06}). Later significant efforts were made for the
identifications of the unknown \int\ X-ray sources, and new CVs were also
revealed \cite{Masetti.etal:08, Masetti.etal:09, Rodriguez.etal:09,
Masetti.etal:10, 2012AstL...38..629K, 2012AstL...38....1L,
Bernandini.etal:12, Masetti.etal:13}. Definitely, it is not possible to
mention all papers devoted to identifications of the {\it INTEGRAL}
sources, but the works which allowed to establish new CVs are listed in
Table\,\ref{tab:int_compil}.

The signature feature of IPs are coherent X-ray pulsations  with periods
ranging from few hundreds to few thousands of seconds. The pulsations reflect
the WD spin periods, and for IPs have to be significantly shorter than the
orbital period. Therefore, if a candidate CV with a hard X-ray spectrum  also
displays coherent periodicities, it can be robustly confirmed as a magnetic
CV. Most of the newly discovered CVs were identified as magnetic systems
using  the {\it XMM-Newton} observatory that allows long uninterrupted
pointings and has high sensitivity in the soft 0.1-10\,keV range (see e.g.
\cite{deMartino.etal:19} and references therein), however, \int\ played
important role in selection of the candidates. Other X-ray instruments such
as {\it Swift}/XRT and {\it RXTE} also helped in establishing the CV nature
of a number of \int\ sources (see, e.g., \cite{Bruns.etal:09,
Butters.etal:08, Butters.etal:11, SDW19}).

Another powerful method to identify CVs is optical spectroscopy. Optical
spectra of such objects are characterised by blue
continua with many emission lines, mainly corresponding to the Balmer
series. Other typical emission features in optical spectra of magnetic CVs
are the HeII ($\lambda 4686$) line and the emission Bowen fluorescence blend
(CIII/NIII spectral lines, seen mainly in the polars). The HeII emission
arises from photoionization of material in the accretion disc and
magnetically confined accretion flow by X-rays while the latter, the Bowen
blend, originates in the irradiated face of the late-type companion star.
The common criterion for IPs is an existence of a significant
HeII$\lambda$4686 emission line with the Equivalent Width (EW) of about
10\AA\ . The ratio of EWs between the HeII and H$_\beta$ lines provides a good
discriminant for IPs, if EW(HeII)/EW(H$_\beta$)\,$>$\,0.5 \cite{vPV:84}.
In turn,  spectral identification of the optical counterpart as an
evolved red giant star (spectral classes KIII - MIII) is a sufficient
base to suggest that the given {\it INTEGRAL} source is a
symbiotic star (see, e.g., \cite{Masetti.etal:08, Fortin.etal:18}).

\section{Models of hard X-ray emission sources in magnetic CVs}
\label{sect:models}



It is commonly accepted now that matter in-falling along the magnetic
field lines forms a strong hydro-dynamical shock above the WD surface. The
post-shock temperature is close to the virial one \cite{Aizu73, FPR76}:

\begin{equation} \label{eq:tsh}
    kT_{\rm sh}=\frac{3}{16}\mu m_{\rm H}  V^2_{\rm ff} \approx 32\, M_1 R_9^{-1} \,\,\,{\rm keV},
\end{equation}

\noindent
where $\mu \approx 0.62$ is the mean molecular weight of plasma with
 solar abundances, $m_{\rm H}$ is the proton mass, $V_{\rm ff}$
is the pre-shock free-fall velocity of the matter, $M_1$ is the WD
mass in solar masses, and $R_9$ is the WD radius in units of
$10^9$\,cm. Here we also assume that the kinetic energy of the ions
converts equally efficient to the thermal energy of ions and
electrons.

As it was mentioned above, the heated plasma in the PSR settles down as a
subsonic flow on the WD surface and cools
mainly due to a free-free and cyclotron emissions \cite{FPR76, LM79}.
The plasma velocity and its temperature are decreasing to values typical
for the WD surface at the bottom of PSR. The height of the PSR
is then determined by the cooling rate of the plasma. This implies that
the settling time in the PSR has to be equal to the cooling time of
the post-shock plasma. Such post-shock structures are sources of the
observed X-ray emission in magnetized CVs.

The observed hard X-ray flux of the PSRs is, therefore, produced by
free-free transitions in the hot post-shock plasma. The PSRs are
optically thin for emerging free-free radiation, so hard X-ray
spectra can be described as an optically thin thermal bremsstrahlung
in the simpler case of a plasma with Hydrogen composition.
The only free parameter in this case is the temperature, which appears
to be an adequate estimate of the post-shock average temperature (see,
however, next subsection). This opens the possibility to estimate the WD
mass as the relation between WD mass and radius are well known
\cite{N72}.
This approach was first suggested and used by Rothschild et al.
(\cite{Rothschild.etal:81}, see also \cite{Ishida91}). It is necessary
to note that these estimates can only be correct for relatively weakly
magnetized objects, i.e. the IPs. The magnetic field strengths at
the WD surface in these objects are $<1-3\times 10^6$\,G
\cite{Warner03,SDW19}, which provide the necessary condition for the
free-free emission as a dominant cooling mechanism in PSRs in comparison
with the cyclotron cooling \cite{LM79}.

The total energy losses due to the cyclotron radiation depend both on the
local plasma temperature and magnetic field strength. Therefore, they are
most significant just after the shock, which leads to reduction of the
maximum PSR temperature and, as a consequence, the observed bremsstrahlung
temperature of the X-ray spectra. This is probably one of the reasons why not
all polars are hard X-ray sources. However, it is important to note that the
free-free cooling  can dominate over the cyclotron cooling for luminous
polars accreting at high rates, because of the high plasma density
\cite{WB96}. Therefore, such objects can be observed as hard X-ray sources.

Soon after the paper \cite{Rothschild.etal:81} it was recognized that a
single bremsstrahlung spectrum is a rather crude approximation for the
emergent spectra of PSRs.  \cite{DO97} suggested to use a number of spectra
of  optically thin plasma, since the optically thin plasma is
multi-temperature. Therefore they used the cooling flow model with the
emission measure power defined by the temperature distribution in the flow,
i.e. $EM \sim (T/T_{\rm sh})^\alpha$. This approach is still used to model
hard X-ray spectra of some intermediate polars  (see, e.g.
\cite{Wada.etal:18}). Fully hydro-dynamical PSR models (see below) can also
be fitted  to this relation (see Table.\,2 in \cite{FBS05} for the comparison
between different PSR models). Another possible approximation  is to assume
that the gas pressure is constant in the PSR \cite{FKR02}. In this case
$\alpha = 0.5$,  and such a model was also used to study X-ray spectra of IPs
(see e.g. \cite{BOH00,Luna.etal:15}).

\begin{figure}
\centering
\includegraphics[angle=0,scale=0.8]{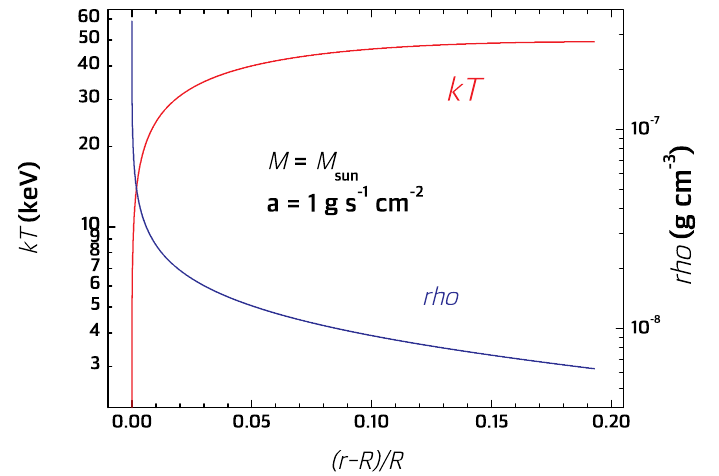}
\includegraphics[angle=0,scale=0.7]{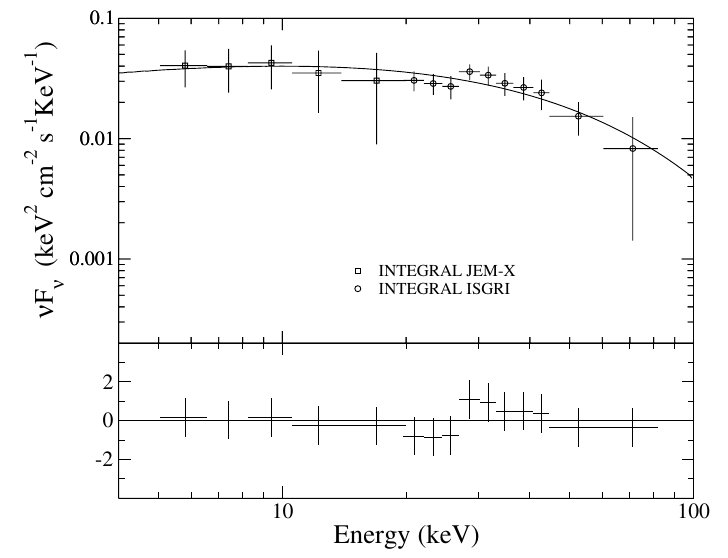}
\caption{\label{fig_psr} Left: Distributions of  temperature (red line) and
density (blue line) in the model with $M = M_{\odot}$ and  local mass
accretion rate $a= 1$~g s$^{-1}$ cm$^{-2}$. Right: The unfolded \int\
JEM-X/ISGRI spectrum of V709\,Cas with the best-fit post-shock model.
Residuals between the data and model are shown below in units of sigma (from
\cite{Falanga.etal:05}).}
\end{figure}

A more correct PSR model can be obtained using a self-consistent
hydrodynamical description. Many works were devoted to this problem, and we
can mention here just a few of them (\cite{Wu.etal:94, WB96,
Cropper.etal:99}). The simplest model describes the optically thin PSR in the
cylindrical geometry cooling only by a thermal radiation without a cyclotron
one. We note that more physical and geometrical details should be included
into the model to make it fully correct (see, e.g. the dipole geometry
importance considered in \cite{Canalle.etal:05}, and the two-temperature
plasma approach studied in \cite{WB96, Saxton.etal:07}). Nevertheless, even
this simplest model is enough to describe adequately X-ray spectra in the
hard energy band. An example of the PSR structure and spectrum of the bright
IP V709\,Cas, measured with \int,  are shown in Fig.\,\ref{fig_psr}.

This approach has been implemented in \cite{SRR05} (see also
\cite{Suleimanov.etal:16}). In the simplest case the model spectrum can be
computed as a sum of the local bremsstrahlung spectra. This is adequate to
describe only the hard continuum above  $\gtrsim$ 10\,keV  since the  soft
part of the observed spectra  is dominated by spectral lines and
photo-recombination continua. Therefore, the sum of the optically thin plasma
spectra over the PSR height rather than multi-temperature bremstrahlung has
to be used in this case. In addition, a complex absorption by intervening
multiple partial covering neutral material further modifies the soft X-ray
spectra of mCVs \cite{Mukai:17}. The detection of the fluorescent Fe $\rm
K_\alpha$ line at 6.4\,keV indicates the presence of reflection from cold
matter, either from the WD or the neutral pre-shock material, which should
contribute in the continuum above 10\,keV. Note that a significant Compton
reflection component was found necessary in fitting high S/N spectra of
bright IPs as observed by {\it NuSTAR} \cite{Mukai.etal:15, Wada.etal:18}.

To avoid the complications associated with the complex nature of the
broadband X-ray spectra of IPs for the WD masses determination, it is
reasonable, therefore, to restrict the modeling to the hard portion of the
observed spectrum   ($E > 15-20$\,keV) \cite{Suleimanov.etal:16,
Hailey.etal:16, SDW19}.

Another complication, which may make WD mass estimates ambiguous, is a finite
size of the magnetosphere which defines the height from which the material
falls onto the surface. The free-fall velocity at the shock depends on the
magnetospheric radius $R_{\rm m}$, which is determined as the inner radius of
the accretion disc disrupted by the WD magnetic field. Obviously in this case
it is possible to determine only a combination of two values: the WD mass $M$
and the relative magnetospheric radius $r_{\rm m}= R_{\rm m}/R$.  As a
consequence, a degeneracy between these two parameters appears. To break this
degeneracy and estimate the WD mass, the magnetospheric radius needs to be
evaluated independently. So far two approaches have been used to that avail.
First, the corotational radius is the natural upper limit for $R_{\rm m}$ and
can be used for the approximate estimate of the latter one. Another way is
that in some cases the frequency of the break in the power spectrum of the
given IPs can be measured, and this frequency is assumed equal to the Kepler
frequency at the magnetospheric radius (see \cite{Revnivtsev.etal:09}). The
implementation of this approach is described in \cite{Suleimanov.etal:16,
SDW19}, and is available as the {\sc XSPEC}
model\footnote{https://heasarc.gsfc.nasa.gov/xanadu/xspec/models/ipolar.html}.
It is necessary to note, however, that the finite magnetospheric radius
affects on the WD mass determination only in the case of the small
magnetosphere size $r_{\rm m}= R_{\rm m}/R < 10$.

The hard X-ray radiation is also observed in non-magnetic CVs, e.g.,
in some dwarf novae during quiescence, and even Nova-like variables
(see Table\,\ref{tab:int_compil}). The hard X-ray radiation in these
systems originates from the optically thin boundary layer between fast
rotating accretion flow  and slow rotating WD \cite{PS79, Tylenda81,
NP93}, or even from the optically thin part of the optically thick
boundary layer \cite{Pat.Ray:85}. Spectra of the optically thin
boundary layers are also well fitted with the cooling flow model
\cite{DO97}. The generally accepted theory of boundary layers does not
yet exist, and we will not describe here any details of different
models.

It is important to note, however, that the maximum bremsstrahlung
temperature should correspond to the Keplerian velocity at the WD
surface, rather than free-fall velocity. As a consequence, the maximum
bremsstrahlung temperature has to be by factor of two smaller for the
optically thin boundary layer in comparison with the post-shock
temperature of a magnetic WD with the same mass.
Therefore, the maximum bremsstrahlung temperature of the hard X-ray
spectra of non-magnetic CVs can be used for WD mass estimations
taking into account the remark above. The hard X-ray radiation of
Symbiotic stars is believed to have similar origin as  in non-magnetic CVs.

\section{White dwarf masses in magnetized CVs according to the {\it \mbox{INTEGRAL}} measurements}
\label{sect:masses}

The first estimate of a WD mass with {\it INTEGRAL} was made by
\cite{Revnivtsev.etal:04a} using the bremsstrahlung temperature obtained from
the approximation of the hard X-ray spectrum of the bright IP V1223\,Sgr.
Subsequently, \int\ observations were used to estimate masses of several WDs
using the PSR model and corresponding spectral model grid \cite{SRR05}. In
particular, IBIS/ISGRI spectra were used together with the {\it RXTE} data to
evaluate the WD mass in V2400\,Oph \cite{Revnivtsev.etal:04b}; a combination
of the IBIS/ISGRI data with JEM-X data was used to determine the WD mass in
V709\,Cas \cite{Falanga.etal:05}. Later this model grid was used to estimate
the WD masses in many IPs (see, e.g., \cite{Brunschweiger.etal:09,
Anzolin.etal:09, Bernandini.etal:12}).

The extended {\it INTEGRAL} observations of CVs allowed to measure
their hard X-ray spectra, which were fitted with the bremsstrahlung and
power-law models \cite{Barlow.etal:06, Landi.etal:09}, but these were
not used for the WD mass determination. We compiled all published WD
mass measurements using the {\it INTEGRAL}
data in Table\,\ref{tab:int_mwd}. In line with the corresponding
temperatures of the bremsstrahlung emission \cite{Barlow.etal:06,
Landi.etal:09}, and for comparison, we also added the WD masses in some
IPs recently determined using {\it NuSTAR} and {\it Swift}/BAT
observations and more sophisticated PSR models \cite{SDW19}.

\begin{figure}
\centering
\includegraphics[angle=0,scale=0.75]{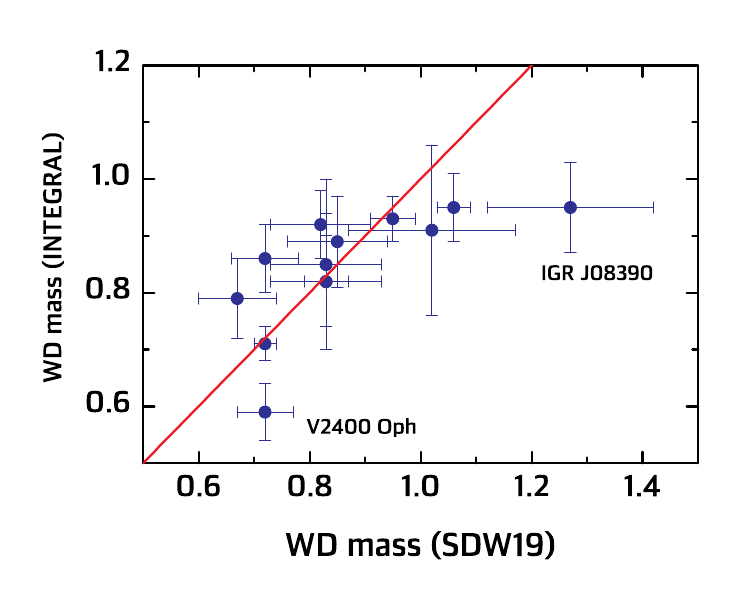}
\includegraphics[angle=0,scale=0.75]{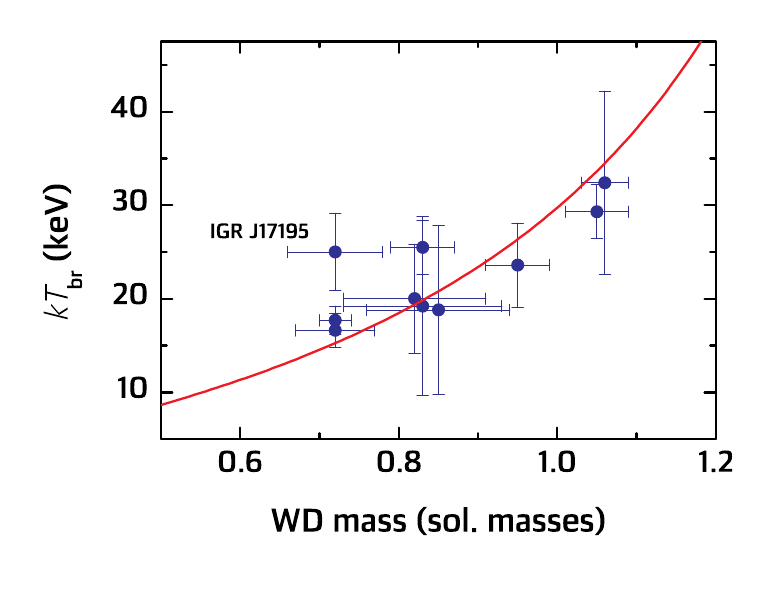}
\caption{\label{fig_mwd}
Left: Comparison of the WD masses of the IPs determined using {\it
INTEGRAL} data with those obtained by \cite{SDW19}. Two outlier
sources are marked. Right: Dependence of bremsstrahlung temperatures
$kT_{\rm br}$ derived from the hard X-ray spectra of some IPs
\cite{Landi.etal:09} on the WD mass in these IPs \cite{SDW19}.
The red curve shows the fit of this dependence with the relation
(\ref{kT_mwd}).
}
\end{figure}

The comparison between the WD masses derived from {\it INTEGRAL}
observations and the values taken from \cite{SDW19} are shown in
Fig.\,\ref{fig_mwd}, left panel. Most of the measurements well agree
one with another within their 1$\sigma$ uncertainties, with only two
IPs, V2400\,Oph and IGR\,J08390-4833, within 3$\sigma$.

As was mentioned above, one of the most straightforward ways to
estimate the WD mass in IPs is to use the single bremsstrahlung
temperature $kT_{\rm br}$ obtained from the simple fitting
of their hard X-ray spectra. The simplest assumption in this approach
is that the bremsstrahlung temperature equals to the maximum
temperature of PSR $kT_{\rm sh}$ (see Eq.\,\ref{eq:tsh}). However, the
averaged bremsstrahlung temperature has to be lower than $kT_{\rm sh}$,
$kT_{\rm br} = A\,kT_{\rm sh}$, where $A<1$. To obtain a real
dependence of $kT_{\rm br}$ on the WD mass we show the bremsstrahlung
temperatures of IPs, which were found in \cite{Landi.etal:09}, as a
function of WD masses from \cite{SDW19} (Fig.\,\ref{fig_mwd}, right).
Note that we consider here only IPs with  a well defined
$kT_{\rm\,br}$, i.e. with the uncertainty on $T_{\rm br}$ smaller than
$kT_{\rm br}/2$. The resulting dependence is well fitted with an
approximate relation

\begin{equation} \label{kT_mwd}
    kT_{\rm br}=A\,kT_{\rm sh} = \frac{A\,32\,M_1}
    {1.364\times(1-0.59\,M_1)}\,~~{\rm keV},
\end{equation}

where $A=0.52\pm0.03$. Here we used a linear fit for the $M-R$ relation
\cite{Suleimanov.etal:16}. The final expression for calculations of the
WD mass using $kT_{\rm br}$ can be written as

\begin{equation} \label{mwd_kT}
    M_1 = \frac{kT_{\rm br}}{A\times23.46+ 0.59\,kT_{\rm br}}.
\end{equation}

Note that this relation is correct for $M_1 < 1.2$ only, because the
linear $M-R$ fit is not valid for larger masses. Nevertheless it
represents a useful way for the quick WD mass estimates using only hard
X-ray data.

Based on the relation (\ref{mwd_kT}) we estimated WD masses in three
systems, the polar IGR\,J14536-5522, and the two IPs, IGR\,J16167-4957
and 2XMMi\,J180438.7-145647) with well determined bremsstrahlung
temperatures. Worth mentioning  is the estimated mass for the AM Her
type system V1432\,Aql (see Table\,\ref{tab:int_mwd}) that is in
remarkable agreement with that derived from {\it RXTE}/PCA data
($M_1\approx$\,0.98) by \cite{Ramsay:00}.

\begin{table}
\caption{Observed and derived parameters of magnetic CVs.
 \label{tab:int_mwd}
 }
{\footnotesize
\begin{center}
\begin{tabular}[c]{ c | l l c c  c c l}
\hline
  &&&&&&& \\
N & 	Name   & Type & $M_1^a$ & $kT_{\rm br}^b$ (keV) &
$kT_{\rm br}^c$ (keV) & $M_1^e$ & Ref.  \\
  &&&&&&& \\
\hline
\hline
  &&&&&&& \\
1 &	V1033 Cas$^*$	& IP & 1.02$\pm$0.15 &  $>$14 &  15.9$\pm$5.1 & 0.91$\pm$0.15 & \cite{Anzolin.etal:09} \\
2 & V709 Cas & IP &  0.83$\pm$0.04 & 25.5$^{+3.1}_{-2.7}$  &    23.3$\pm$2.2 & 0.82$^{+0.12}_{-0.25}$ &  \cite{Falanga.etal:05} \\
3 &  V515 And  &   IP &  0.67$\pm$0.07 & & & 0.79$\pm$0.07 & \cite{Bernandini.etal:12}\\
4 &	GK Per	& IP &  0.79$\pm$0.01  & 43.7$^{+128.8}_{-23.4}$ &  28.7$\pm$15.7 & &\\
5 & BY Cam	&  AM &  &     36.7$^{+230.3}_{-22.7}$  &        14.8$\pm$5.9 & {\bf $>$\,0.72} & \\
6  & MU Cam  &  IP & 0.67$\pm$0.08 &  7.0$^{+8.9}_{-3.8}$  &  8.1$\pm$4.7 &  & \\
7 & IGR J08390-4833$^*$  & IP &    1.27$\pm$0.15 & & & 0.95$\pm$0.08 & \cite{Bernandini.etal:12}\\
8 &	V834 Cen	& AM &      &     18.7$^{+61.0}_{-10.6}$    &    19.5$\pm$7.8 & {\bf $>$\,0.5}& \\
9 & IGR J14536-5522$^*$ &  IP &  &  14.0$^{+5.4}_{-3.7}$ &   11.1$\pm$4.6 & {\bf 0.68$\pm$0.13} & \\
10 & IGR J15094-6649$^*$ &  IP & 0.85$\pm$0.09 & 18.8$^{+11.5}_{-6.5}$ &  13.8$\pm$5.1  & 0.89$\pm$0.08 & \cite{Bernandini.etal:12} \\
11 &  NY Lup	& IP & 1.05$\pm$0.04  &   29.3$^{+3.1}_{-2.7}$     &   27.1$\pm$ 2.2 & & \\
12 &  IGR J16167-4957$^*$  &  IP &    &  17.3$^{+3.3}_{-2.7}$ & 13.2$\pm$2.6 & {\bf 0.77$\pm$0.07} &\\
13 & 	IGR J16500-3307$^*$ & IP &   0.82$\pm$0.09  &  20.0$^{+6.9}_{-4.7}$ & &  0.92$\pm$0.06 & \cite{Bernandini.etal:12}\\
14 & 	IGR J16547-1916$^*$ &	 IP & 0.83$\pm$0.10 & & & 0.85$\pm$0.15 &\cite{Lutovinov.etal:10}\\
15 & V2400 Oph	& IP & 0.72$\pm$0.05   &  16.6$^{+1.9}_{-1.7}$     & 18.6$\pm$1.4 & 0.59$\pm$0.05 &  \cite{Revnivtsev.etal:04b}\\
16 & IGR J17195-4100$^*$ & IP &  0.72$\pm$0.06 & 25.0$^{+4.6}_{-3.6}$ &  27.0$\pm$4.4 & 0.86$\pm$0.06 & \cite{Bernandini.etal:12}\\
17 & V2731 Oph$^*$ & IP &  1.06$\pm$0.03  &   32.4$^{+11.5}_{-8.1}$ &  26.7$\pm$4.4 & 0.89 -1.02 &
\cite{deMartino.etal:08}\\
18 & 	V2487 Oph  & IP? &      & 55.6$^{+76.4}_{-24.5}$   &   25.5$\pm$8.6 & {\bf $>$\,1.0}&\\
19 &  2XMMi J180438.7-145647 & IP & & & (23.3$^{+8.84}_{-5.9})^d$ &  {\bf 0.90$^{+0.13}_{-0.11}$}
& \\
20 & IGR J18308-1232$^*$ & IP &  & & & 0.85$\pm$0.06 & \cite{Bernandini.etal:12}\\
21 & V1223 Sgr & IP & 0.72$\pm$0.02 & 17.7$^{+1.6}_{-1.4}$  &  18.8$\pm$1.2 & 0.71$\pm$0.03 &\cite{Revnivtsev.etal:04a}\\
22 & V1432 Aql &  AM &  & 24.4$^{+15.3}_{-8.1}$  &  25.4$\pm$7.0 & {\bf 0.92$\pm$0.18} &\\
23 & V2069 Cyg & IP & 0.83$\pm$0.10 & 19.2$^{+12.3}_{-6.8}$&  35.7$\pm$16.8 & 0.82$\pm$0.08 & \cite{Bernandini.etal:12} \\
24 & IGR J21335+5105$^*$ &  IP & 0.95$\pm$0.04 & 23.6$^{+5.0}_{-4.0}$ &  23.8$\pm$4.3 & 0.93$\pm$0.04 & \cite{Anzolin.etal:09}\\
25 & FO Aqr	&  IP &  0.57$\pm$0.03 & 29.7$^{+70.1}_{-16.6}$ & &  & \\ \hline
\end{tabular}
\end{center}
Note: $^*$ -- discovered by {\it INTEGRAL}, IP -- an intermediate
polar, AM -- a polar, type AM Her, $a$ -- the WD masses from
\cite{SDW19}, $b$ -- $kT_{\rm br}$ from \cite{Landi.etal:09},
$c$ -- $kT_{\rm br}$ from \cite{Barlow.etal:06}, $d$ -- $kT_{\rm
br}$ taken from \cite{Middleton.etal:12}, $e$ -- the WD masses obtained
using INTEGRAL observations. The corresponding references are shown in
the last column. The boldfaced WD masses given without references
were obtained in this work using relation (\ref{mwd_kT}), and the
bremsstrahlung temperatures measured in \cite{Landi.etal:09,
Barlow.etal:06, Middleton.etal:12}.
}
\end{table}

It is important to note that for some objects the bremsstrahlung temperatures
$kT_{\rm br}$ reported in \cite{Landi.etal:09} have large uncertainties,
despite the relatively high brightness of these sources
(Table\,\ref{tab:int_mwd}). A possible reason of this is that their
temperatures were determined using the spectra accumulated over a long time
period (over five years). This implies that the large $kT_{\rm br}$
uncertainties for the brighter sources may be associated with the intrinsic
variability of the spectrum related to mass accretion rate variations, which,
in turn, lead to the variations of the magnetospheric radius. For these
objects only lower limits on their WD masses are reported in
Table\,\ref{tab:int_mwd}, if any other WD mass determinations are absent.
Thus, these systems require additional dedicated investigations to draw
conclusions regarding the masses of their WD primaries.

\section{Symbiotic stars}
\label{sect:SyS}

Symbiotic stars, long-period binary systems where either a white dwarf
or a neutron star accretes from the wind of a red giant companion, have
been known as X-ray sources since the 70's (\cite{Lewin.etal:71,
Anderson.etal:81, Cordova.etal:81, Allen:81}). The characteristics of
the X-ray emission from symbiotic stars were compiled by
\cite{Muerset.etal:97} using {\emph ROSAT} data.

Symbiotic stars were classified according to the following scheme (see
\cite{Muerset.etal:97} and following development by \cite{Luna.etal:13}):

\begin{itemize}
\item[$\alpha$:] \emph{supersoft X-ray sources.} Most of their X-ray
radiation is emitted below $\approx 0.4$\,keV. They are supposed to be
white dwarfs with a quasi-steady shell burning on their surface;
\item[$\beta$:] \emph{soft X-ray sources.} Their X-ray spectra extend
up to $\approx 2.4$\,keV (i.e., the upper bound of the \emph{ROSAT}
energy range). The X-ray photons are likely produced by the collisions
of the winds from the white dwarf and the red giant;
\item[$\gamma$:] \emph{symbiotic stars with accreting neutron stars (or X-ray symbiotic stars).} They have the hard X-ray emission ($E \gtrsim
2.4$\,keV), mostly from the optically thick Comptonizing plasma;
\item[$\delta$:] \emph{hard X-ray sources with high absorption.} They
show a strong thermal emission above $E\approx 2.4$\,keV, likely
originating in the accretion disc boundary layer;
\item[$\beta/\delta$:] \emph{symbiotics with soft and hard thermal
components.} They have properties in common with the subclasses $\beta$
and $\delta$, and the mechanisms producing both spectral components are
supposed to be the same of those proposed for $\beta$ and $\delta$
sources.
\end{itemize}

Only two systems with neutron stars as accreting object were present in
the compilation by \cite{Muerset.etal:97}: GX\,1+4 and
Hen\,3-1591.  In the modern era of instruments sensitive to energies of
more than 10 keV such as {\it INTEGRAL} and $Swift$/BAT, about 10 more
similar systems have been discovered.  It is believed that their X-ray
emission arises from an optically thick Comptonizing plasma with no
emission lines \cite[see][]{Marcu.etal:11}. These systems are out of
scope of the current review and will be reviewed separately in this volume.

\begin{figure}
\centering
\includegraphics[angle=-90,trim={0 0 0 0},width=0.95\columnwidth]{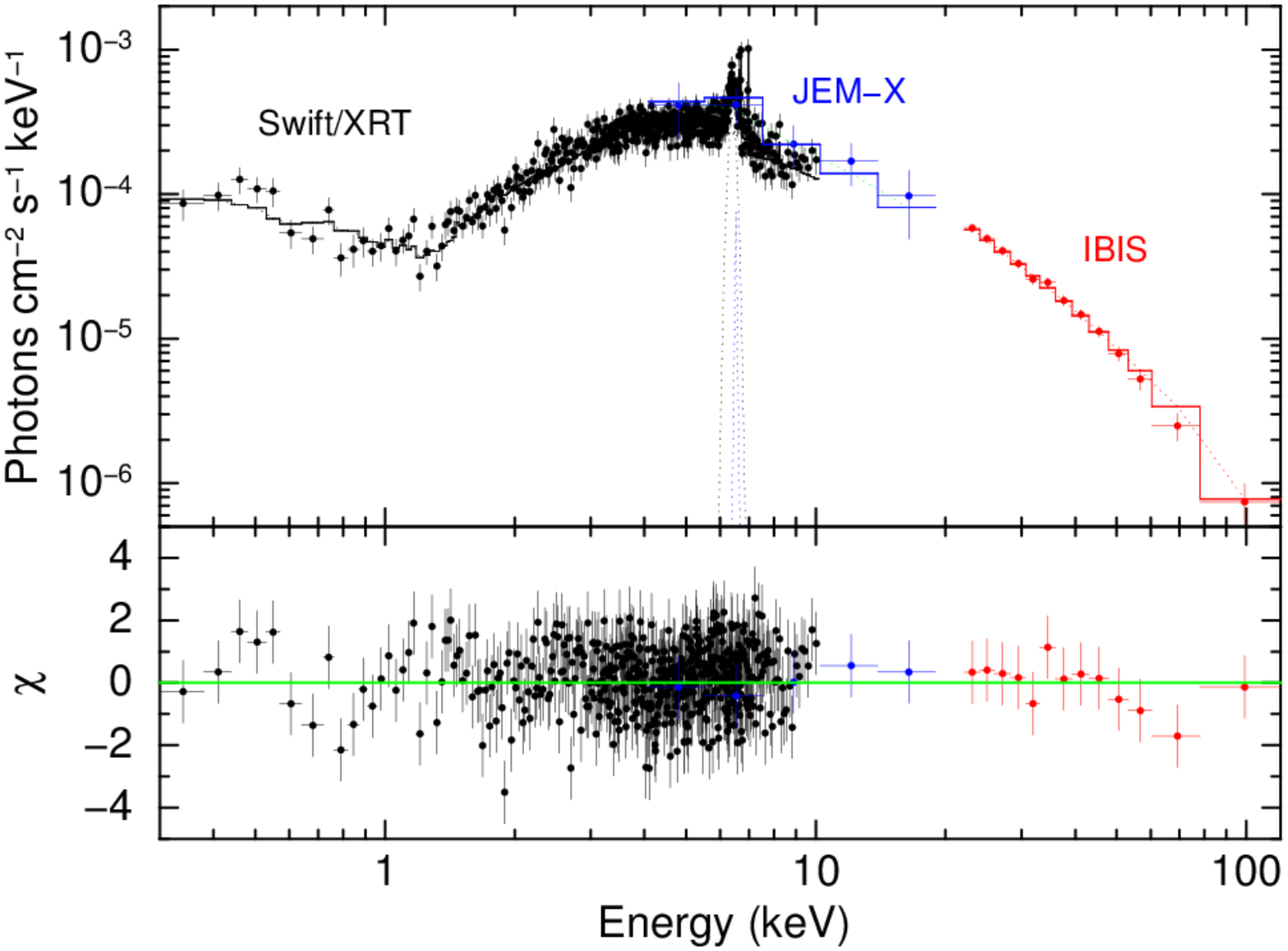}
\caption{\label{fig_ducci}Combined {\em Swift}/XRT (black), JEM-X (blue),
IBIS/ISGRI (red) unfolded spectra of RT\,Cru/IGR\,J12349-6434. The \int\
spectrum was constructed accumulating data from 2003 January 29 and 2014
December 20, while for {\em Swift}/XRT the spectrum was built accumulating
data from Aug 20, 2005 to Dec 24, 2012 \cite[see][for
details]{Ducci.etal:16}.}
\end{figure}

The vast majority of symbiotic systems host, however, accreting WDs, making
them long-period relatives of CVs. After the aforementioned compilation
by \cite{Muerset.etal:97}, \cite{Luna.etal:13} reported the first
detection of the X-ray emission from eight more symbiotic systems using
$Swift$/XRT, thus bringing the total number of known symbiotic systems
as X-ray sources to 33 (see also \cite{Mukai:17}).

Four systems stood out from this sample, RT\,Cru, T\,CrB, CD\,-57\,3057
and CH\,Cyg, because of their unprecedented high energy emission. The
prototype of them is RT\,Cru, which was first detected with {\it
INTEGRAL}/IBIS in 2003-2004 with the name of IGR\,J12349-6434 at a
$\sim3$ mCrab level in the 20-60 keV energy band
\cite{Chernyakova.etal:05}. Later, {\it INTEGRAL} detected RT\,Cru
again in 2012 with a flux of $\sim13$ mCrab in the 18-40 keV band
\cite{Sguera.etal:12}, and in 2015 at the level of $\sim6$ mCrab in the
22-60 keV band \cite{Sguera.etal:15}. The joint {\em Swift}/XRT+{\em
INTEGRAL} JEM-X/IBIS/ISGRI spectrum from RT Cru is shown in
Fig.\,\ref{fig_ducci}. The system CD\,-57 3057 was also registered with
{\it INTEGRAL} during the Crux Galactic arm survey as the IGR\,J10109-5746
source \cite{Revnivtsev.etal:06}. Moreover, \int\ discovered several new
hard X-ray sources, which are considered symbiotic
star candidates. The full list of such sources can be found in \int\
catalogues, but here we can mention the most promising objects:
IGR\,J15293-5609 (=CXOU\,J152929.3-561213) \cite{2012ApJ...754..145T},
IGR\,J17164-3803 \cite{2017MNRAS.465.1563R} and IGR\,J17463-2854
\cite{2015AstL...41..394K}.

Because of their hard X-ray emission, these objects were dubbed as
$\delta$-type X-ray sources in \cite{Luna.etal:13}. This hard X-ray emission
appears to be thermal, and the X-ray spectrum can be modeled with a highly
absorbed cooling flow model (see Sect. \ref{sect:models}) with maximum
temperatures of about 50 keV \cite{Smith.etal:08,Luna.etal:07,Eze.etal:10}.
Such hard X-ray emission most likely originates in the most internal region
of the accretion disk, the boundary layer, which is optically thin to its own
radiation if the accretion rate is low, $\dot{M} \lesssim 3\times 10^{-10}$
M$_{\odot}$ yr$^{-1}$ for a 1\,M$_\odot$ white dwarf
\cite{Popham.etal:95,Suleimanov.etal:14}. Conversely, the hard X-rays could
arise in the accretion column of a magnetic WD, similarly to IPs, although
the WD period has not been detected yet in any system  \cite{Ducci.etal:16}.

\begin{figure}[]
\centering
\includegraphics[angle=0,scale=0.5]{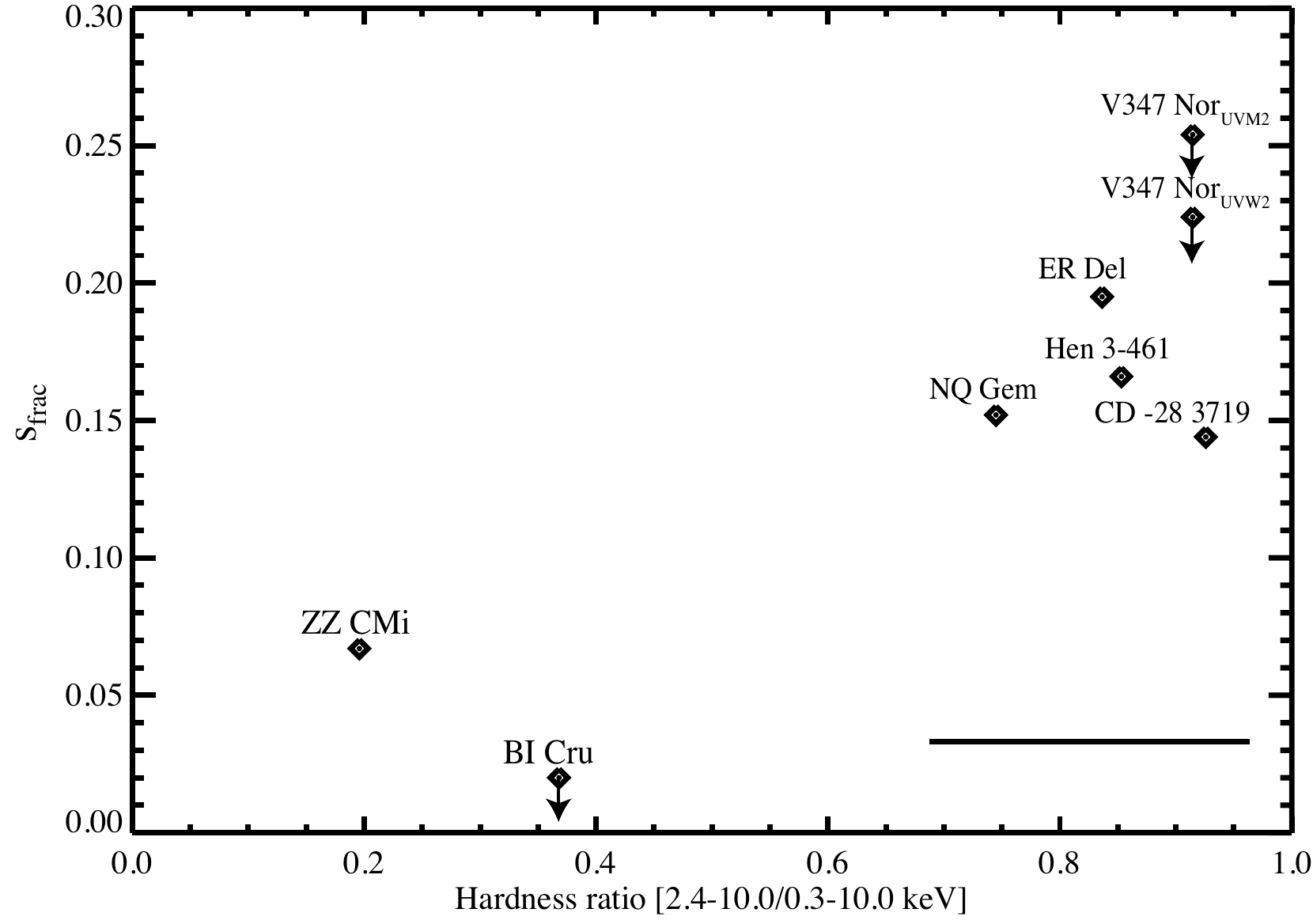}
\caption{\label{fig7_luna} Fractional rms amplitude of flickering
($S_{frac}$) vs. hardness ratio in X-rays. Objects with harder X-ray emission
tend to have more intense UV flickering.  Downward arrows indicate upper
limits. The average error bar is shown at the bottom-right corner
\cite[from][]{Luna.etal:13}}.
\end{figure}

Nevertheless, most of the accretion-powered, hard X-rays emitting, symbiotic
systems are too weak to be detected with the current instruments ($L_{X} \leq
10^{34}$ ergs\,s$^{-1}$). However, Luna et al. \cite{Luna.etal:13} found that
the amplitude of the flickering in the UV band, with time scales of minutes
to hours, (see Fig. \ref{fig7_luna}) is greater in those systems with harder
X-ray emission, while sources with low-amplitude UV flickering tend to have
relatively little emission above 2 keV. This provides a tool to search for
accretion-powered systems while hard X-ray observation would be unfeasible
long.

As in the case of IPs, the post-shock maximum temperature can be used to
determine the WD mass if the X-rays arise in a completely optically thin
boundary layer, otherwise, the maximum temperature provides a lower limit for
the WD mass. The best-studied systems, RT~Cru and T~CrB, show strong,
long-term variability in their X-ray light curves, which points to changes in
the accretion rate and subsequently in the optical depth of the boundary
layer.

\begin{figure}
\centering
\includegraphics[angle=0,scale=0.5]{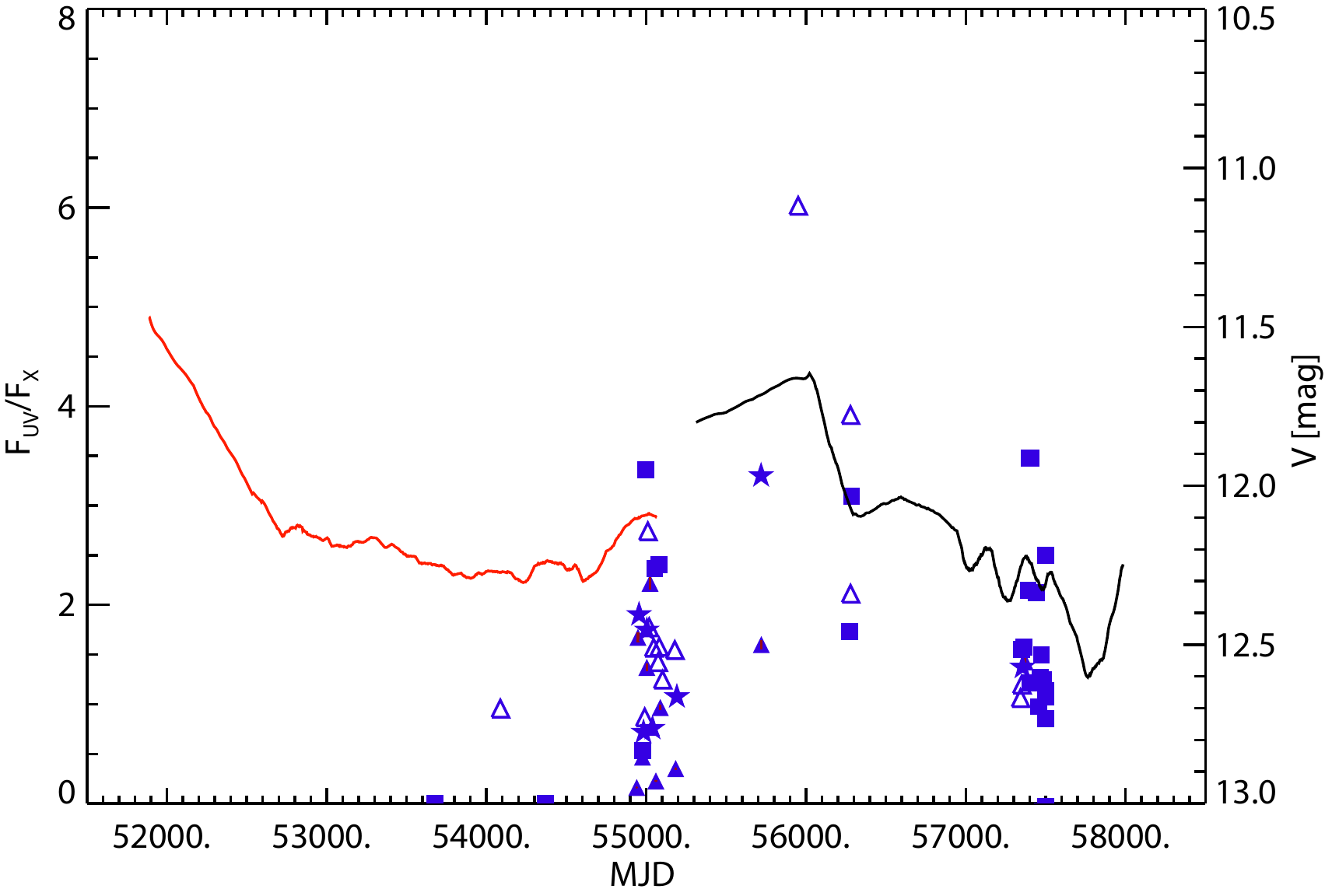}
\caption{\label{fig_rtcru1} ({\it Left-hand Y-axis}) Ratio of
reddening-corrected {\it Swift}/UVOT magnitudes (U: filled triangle;
W1: star; M2: open triangle; W2: square) over unabsorbed X-ray flux
in the 0.3-80 keV band for each {\it Swift} observation of RT~Cru. ({\it
Right-hand Y-axis}) Over-plotted are the ASAS (red solid line) and AAVSO
(black solid line) V-band light curve, smoothed for clarity
\cite[][]{Luna.etal:18a}.}
\end{figure}

In the standard accretion disk theory, the total accretion luminosity is
divided in equal portions between the Keplerian portion of the accretion disk
and the boundary layer. The Keplerian accretion disk radiates mostly in the
UV and optical regimes while the boundary layer does it in X-rays. Thus, the
ratio of $L_{UV}$/$L_{X}$ should be equal to one if the boundary layer is
completely optically thin. During the high-accretion-rate episodes, such as
those in dwarf novae in outburst, or persistent high-accretion regimes as in
novae-like, the boundary layer often becomes optically thick to its own
radiation, and the peak of its emission shifts toward longer wavelengths,
causing the $L_{UV}$/$L_{X}$ ratio to be significantly greater than one.

Multiwavelength, UV and X-rays light curves obtained with $Swift$/XRT and
UVOT can be used to trace the optical depth of the boundary layer.
\cite{Luna.etal:18a}  used data from the $Swift$/XRT and UVOT simultaneously
and found that, in RT~Cru, the $L_{UV}$/$L_{X}$ is greater than one during
the optical maximum while it is closer to one during the optical minimum
(Fig. \ref{fig_rtcru1}). This behavior might be related with the putative
orbital period of about 4000 days found in the optical and $Swift$/BAT light
curves.

Another interesting case is that of T\,CrB, a symbiotic recurrent nova with a
recurrence period of about 80 years. The last thermonuclear outburst happened
in 1946. Given that it is the closest recurrent nova, it will certainly
produce copious hard-X-rays-to-$\gamma$ rays during the next thermonuclear
outburst that could be detected with \int.

Since 2014, T\,CrB entered in what has been called a "super-active" state
\cite{Munari.etal:16}, in which the optical brightness increased by about 1
magnitude, the 15-50 keV flux almost vanished, the UV flux increased by a
factor of about 40 and a new soft X-ray (0.3-1 keV) black body type component
emerged. This phenomenology is akin to what is observed during the dwarf nova
outbursts, in which the most internal region of the accretion disk changes
its optical depth in response to an increase in the accretion rate. However,
given the large size of the accretion disk in T\,CrB, these changes happen in
a far longer time scale.

\section{Luminosity function of CVs}
\label{sect:LumF}

The determination of the space densities of different classes of CVs in the
Galaxy is essential to constrain models of CV formation and evolution.
Although fairly large catalogues of CVs, mainly selected in the optical band,
have been available for a long time (see \citep{Ritter.Kolb:04} and
references therein), they were not well suited for determining CV space
densities because of the strong and poorly understood selection effects
(\cite{Patterson:84}, see also \cite{2015ApJ...809...80G}). A breakthrough
occurred when the \emph{ROSAT} observatory conducted an all-sky soft X-ray
survey and provided a well-defined, flux limited sample of $\sim$50~CVs
(mainly identified among the X-ray sources from the \emph{ROSAT} Bright
Survey at high Galactic latitudes, $|b|>30^\circ$), which has then been used
to estimate the space density of non-magnetic and magnetic CVs
\citep{Schwope.etal:02,Pretorius.etal:12,Pretorius.etal:13}. Recently, an
additional small sub-sample of CVs was identified among the X-ray sources
detected in deeper pointed observations with \emph{ROSAT}, which made it
possible to extend the soft X-ray luminosity function of CVs down to
$10^{29}$\,erg\,s$^{-1}$ \citep{Burenin.etal:16}. However, the \emph{ROSAT}
survey was limited to energies below 2\,keV, whereas, as discussed in depth
in this review, some classes of CVs such as IPs, emit a large or even
dominant fraction of their bolometric luminosity at higher energies.
Therefore, in order to conduct a reliable census of such objects it is highly
desirable to select them in the hard X-ray band. The first representative
sample of CVs selected in a moderately hard X-ray band, 3--20\,keV, was
obtained by \citep{Sazonov.etal:06} based on the \emph{Rossi XTE} Slew Survey
(XSS) \citep{Revnivtsev.etal:2004}. This sample consisted of 24 known CVs,
including 4 non-magnetic ones (dwarf novae), 19 magnetic ones (6 polars and
13 intermediate polars) and 1 symbiotic star. For all but one of these
objects, distance estimates were available (including several accurate
parallax measurements), with most of them located within 500\,pc from the
Sun. This sample was used to construct the 3--20\,keV luminosity function of
CVs spanning three decades in luminosity (from $10^{31}$ to
$10^{34}$\,erg\,s$^{-1}$) and to measure the total space density and total
X-ray luminosity density (per unit stellar mass) of such objects: $(4.8\pm
1.6)\times 10^{-7}$\,pc$^{-3}$ and $(2.4\pm 0.6)\times
10^{27}$\,erg\,s$^{-1}$\,$M_\odot^{-1}$ (3--20\,keV), respectively. A
relevant study was later performed by \citep{Byckling.etal:10}, where
\emph{Suzaku}, \emph{XMM-Newton} and \emph{ASCA} data were used to construct
a 2--10\,keV luminosity function of dwarf novae and it was shown that this
subclass of CVs contributes $\sim$16\% to the total local X-ray luminosity
density of CVs.

Afterwards, \citep{Revnivtsev.etal:08} applied the same approach to the data
of \emph{INTEGRAL} observations and for the first time constructed a sample
of CVs selected in a truly hard X-ray band, 17--60\,keV. To this end, they
used the catalogue of sources detected by the IBIS instrument over the whole
sky during the first 3.5 years of the \emph{INTEGRAL} mission
\citep{Krivonos.etal:07b}. The sample consisted of 17 CVs: 15 intermediate
polars (note that the JEM-X telescope aboard \emph{INTEGRAL} has detected 3
of these objects also in the 5--10 and 10--25\,keV energy bands,
\citep{Grebenev.etal:15}), one dwarf nova, and one classical nova (V2487~Oph,
which also could be a magnetic CV of the IP type \citep{Hernanz14}). As in
previous studies, a serious problem was posed by the highly uncertain
distances to some of the studied objects. If the spectral type of the
secondary star and the apparent brightness of a CV are known, one can usually
attempt to estimate its distance. However, in the case of IPs, the accuracy
of this method is significantly affected by the fact that the optical light
is mainly produced by the accretion disc rather than the secondary star.
Detailed spectroscopic information allows one in some cases to determine the
contribution of the light of the companion star to the optical brightness of
the binary, but this information was not available for all CVs in the
\emph{INTEGRAL} sample. To overcome these difficulties, the authors took the
following novel approach: they measured the correlation between the orbital
period and the hard X-ray luminosity for the six CVs with known distances
from the sample and used the derived empirical, but physically motivated
dependence to predict the hard X-ray luminosities and hence distances for the
remaining objects.

\begin{figure}
\centering
\includegraphics[scale=0.5,trim={0 170 0 100},clip]{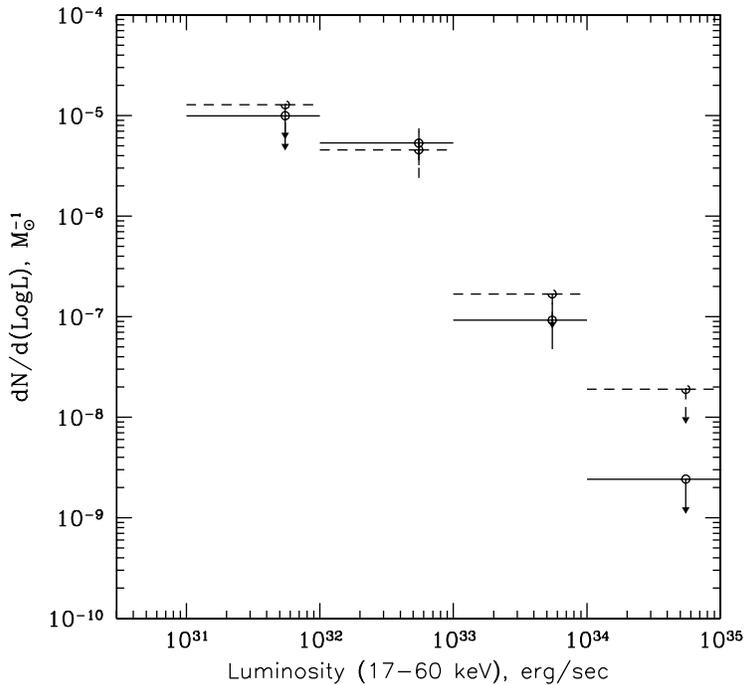}
\caption{Luminosity function of CVs detected by \emph{INTEGRAL} over the
whole sky (solid crosses). The dashed crosses show the luminosity function
obtained excluding the Galactic plane region ($|b| < 5^\circ$), where there
remain unidentified \emph{INTEGRAL} sources. From \citep{Revnivtsev.etal:08}.
}
\label{fig_lumfunc}
\end{figure}

The resulting \emph{INTEGRAL} sample, despite its relatively small size,
effectively covered three orders of magnitude in luminosity,
$10^{32}$--$10^{35}$\,erg\,s$^{-1}$ (in addition, the absence of sources in
the $10^{31}$--$10^{32}$\,erg\,s$^{-1}$ provided an interesting upper limit
on the space density of such objects), which made it possible to construct
the luminosity function of CVs in the hard X-ray band for the first time
(Fig.~\ref{fig_lumfunc}). The integrated space density of CVs with hard X-ray
luminosities above $10^{32}$\,erg\,s$^{-1}$ proved to be $(1.5\pm 0.6)\times
10^{-7}$\,pc$^{-3}$ near the Sun, or equivalently $(3.8\pm 1.5)\times
10^{-6}$\,$M_\odot^{-1}$ (taking into account the local stellar density
$\sim$0.04\,$M_\odot$\,pc$^{-3}$). The total X-ray luminosity density of such
objects was found to be $(1.3\pm 0.3)\times
10^{27}$\,erg\,s$^{-1}$\,$M_\odot^{-1}$ (17--60\,keV). Finally,  it proved
possible to estimate the exponential scale height of CVs (primarily IPs) in
the Galactic disc at $130^{+93}_{-46}$\,pc.

A similar study has been subsequently carried out using the data of the
all-sky hard X-ray survey conducted by the BAT instrument aboard the
\emph{Swift} observatory \citep{Pretorius.etal:2014}. The authors imposed
selection cuts in flux (above $2.5\times 10^{-11}$\,erg\,cm$^{-2}$\,s$^{-1}$
at 14--195\,keV) and Galactic latitude ($|b|>5^\circ$) and obtained a sample
of 19 CVs, including 15 IPs. The latter sub-sample was used to obtain an
estimate of the local space density of IPs with hard X-ray (14--195\,keV)
luminosities above $10^{32}$\,erg\,s$^{-1}$: $1^{+1}_{-0.5}\times
10^{-7}$\,pc$^{-3}$, in excellent agreement with the \emph{INTEGRAL}
measurement.

The recent second release (DR2) of \emph{Gaia} parallaxes
\cite{2018A&A...616A...1G} now offers the opportunity to assess the true
space densities and determine more accurate luminosity function. A first work
using the shallow flux-limits of the 70-month {\it Swift}/BAT sample by
\citep{Pretorius.etal:2014} and \emph{Gaia} parallaxes, the IP space density
is found to be lower than previously estimated with an upper limit of $\rm
<1.3\times10^{-7}\,pc^{-3}$ although consistent within errors with previous
determinations \cite{2018A&A...619A..62S}. A lower CV space density than
predicted by most binary population synthesis models has been recently
confirmed by \cite{2019arXiv190713152P} using a volume limited sample of 42
CVs within 150\,pc. They also find that the fraction of mCVs is very large
$\sim 33\%$. It is therefore crucial to extend the study to the larger
fraction of mCVs detected so far. In  this respect it is worth mentioning the
successful launch of the Russian-German \emph{Spectr-RG} mission in summer
2019. The eROSITA and ART-XC telescopes on board of this mission
\cite{2012arXiv1209.3114M,2018SPIE10699E..1YP} will conduct the most
sensitive X-ray survey in the 2-11\,keV band to date, and thus promise the
discovery of thousands of faint sources pushing the limits of current X-ray
luminosity function and assessing the true CV space densities.

\section{Galactic Ridge X-ray Emission}
\label{sect:ridg}

There are two major large-scale extended diffuse features in the X-ray sky
(above 2 keV): the highly isotropic cosmic X-ray background (CXB) and an
emission concentrated towards the Galactic plane -- the Galactic ridge X-ray
emission (GRXE) \citep{Worrall.etal:82,Warwick.etal:85}. While it became
clear long ago that the CXB is a superposition of numerous active galactic
nuclei, the origin of the GRXE remained a puzzle for a long time.

The GRXE energy spectrum contains a number of emission lines of highly
ionized heavy elements indicating that the radiation originates in an
optically thin plasma with a temperature $\sim 10^8$\,K
\citep{Koyama.etal:86,Tanaka:02}. The total GRXE luminosity is
$\sim$(1--2)$\times 10^{38}$\,erg\,s$^{-1}$ \citep{Valinia.Marshall:98}.
Historically, two points of view were confronted: (i) the GRXE is the
superposition of point-like sources or (ii) it is truly diffuse emission.
Each of these hypotheses had its problems. The proponents of the former
usually tried to account for the observed GRXE luminosity in terms of the
expected contributions from different known classes of faint ($L_X<
10^{34}$\,erg\,s$^{-1}$) point sources, but the space densities involved in
this estimation were not known accurately enough to draw firm conclusions
\citep{Worrall.Marshall:83,Ottmann.Schmitt:92,Mukai.Shiokawa:93}. The main
difficulty with the second scenario was to explain how to keep the hot plasma
within the Galaxy or, if it is outflowing, to find the source of energy
maintaining this strong wind \citep{Tanaka:02}.

In 2006, \citep{Revnivtsev.etal:06b} accomplished a crucial achievement by
demonstrating that the GXRE surface brightness, as mapped in the 3--20\,keV
energy band by \emph{RXTE}, closely traces the near-infrared (NIR) surface
brightness of the Milky Way and hence the distribution of stars throughout
the Galaxy. This strongly suggested that the GXRE consists of point-like
stellar-type sources. The measured X-ray/NIR correlation implied that the
cumulative specific X-ray emissivity of these unresolved X-ray sources is
$(3.5\pm 0.5)\times 10^{27}$\,erg\,s$^{-1} M_\odot^{-1}$ (3--20\,keV).

The same team \citep{Sazonov.etal:06} used the \emph{RXTE} data to construct
the luminosity function of CVs (see Sect. 6) and coronally active binary
stars (ABs, of RS CVn and other types), in the Solar neighbourhood, which
allowed them to determine the integrated 3--20~keV luminosity of such faint
($L_X<10^{34}$\,erg\,s$^{-1}$) X-ray sources per unit stellar mass: $(5.3\pm
1.5)\times 10^{27} M_\odot^{-1}$\,erg\,s$^{-1}$. Remarkably, this locally
measured quantity proved to be consistent, within the uncertainties, with the
X-ray production rate required to explain all of the GRXE (see above).
Furthermore, the spatial distribution of CVs and ABs is expected to trace the
overall distribution of stars in the Galaxy. Therefore, a consistent picture
emerged that the bulk of the GRXE is made up by ABs and CVs. This conclusion
has later received a spectacular confirmation from the direct resolution of
$\sim 80$\% of the GRXE at 6--8\,keV into discrete sources in ultra-deep
\emph{Chandra} observations of a small region near the Galactic Centre
\citep{Revnivtsev.etal:09b,Revnivtsev.etal:11}.

If the GRXE indeed represents the the sum of the emissions from ABs and CVs,
then also the shape of its energy spectrum must be a superposition of
representative spectra of these classes of objects. The hard X-ray spectrum
has always been a major issue in the study of the GRXE. In particular, if it
had a power-law shape, it would imply that non-thermal phenomena, such as
cosmic-ray induced emission \citep{Skibo.etal:97}, are involved in the
formation of the GRXE. The IBIS telescope aboard \emph{INTEGRAL} has for the
first time combined a large field of view with moderate angular resolution,
making it possible to measure the GRXE spectrum by collecting a significant
flux of Galactic "diffuse" hard X-rays and separating out the contribution of
bright point sources.

\begin{figure}
\centering
\includegraphics[width=0.9\textwidth,viewport=150 150 700 400]{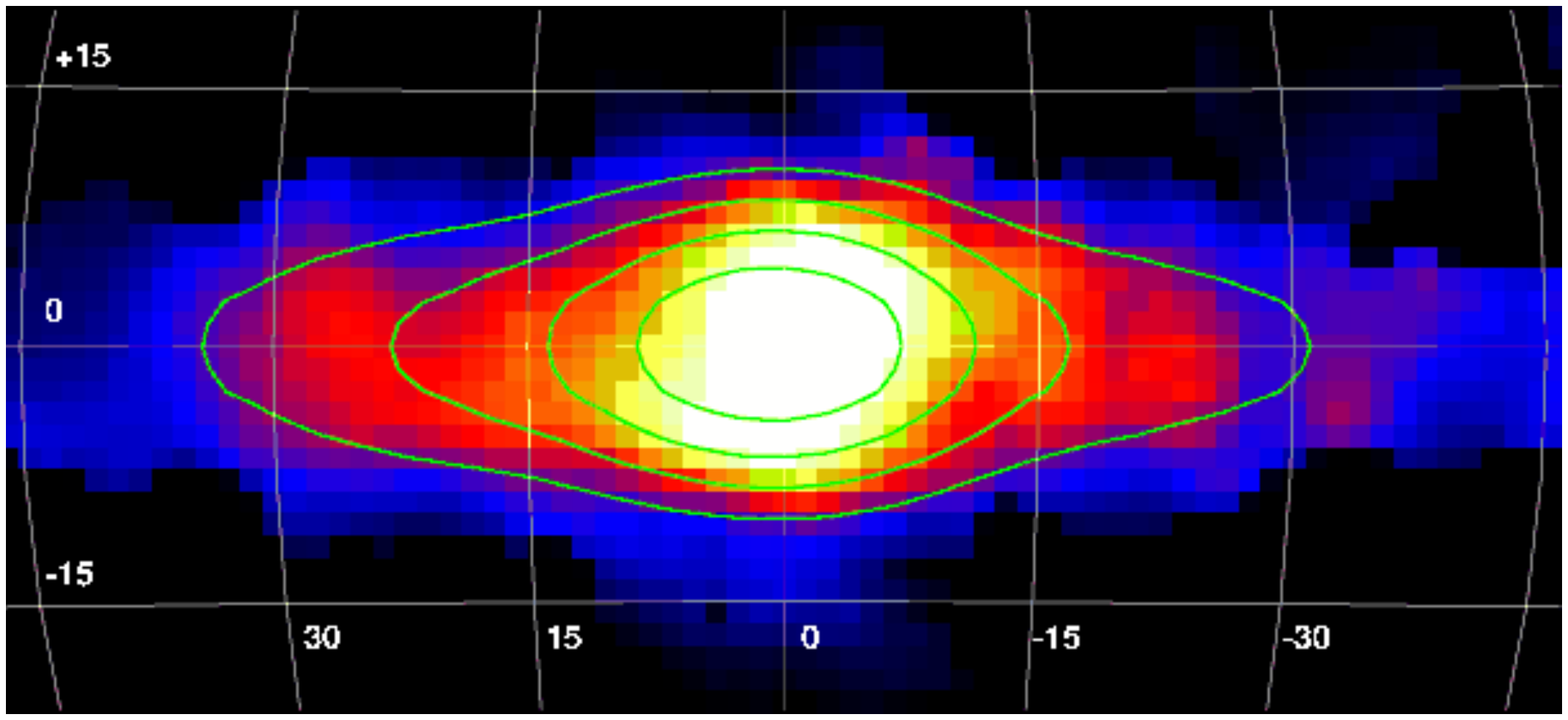}
\caption{Map of the unresolved Galactic hard X-ray (17--60\,keV) emission
obtained with \emph{INTEGRAL}/IBIS, in comparison with the NIR map (contours)
obtained by \emph{COBE}/DIRBE. From \citep{Krivonos.etal:07}. }
\label{fig_grxe_map}
\end{figure}

\begin{figure}
\centering
\includegraphics[width=0.5\textwidth,viewport=175 200 450 525]{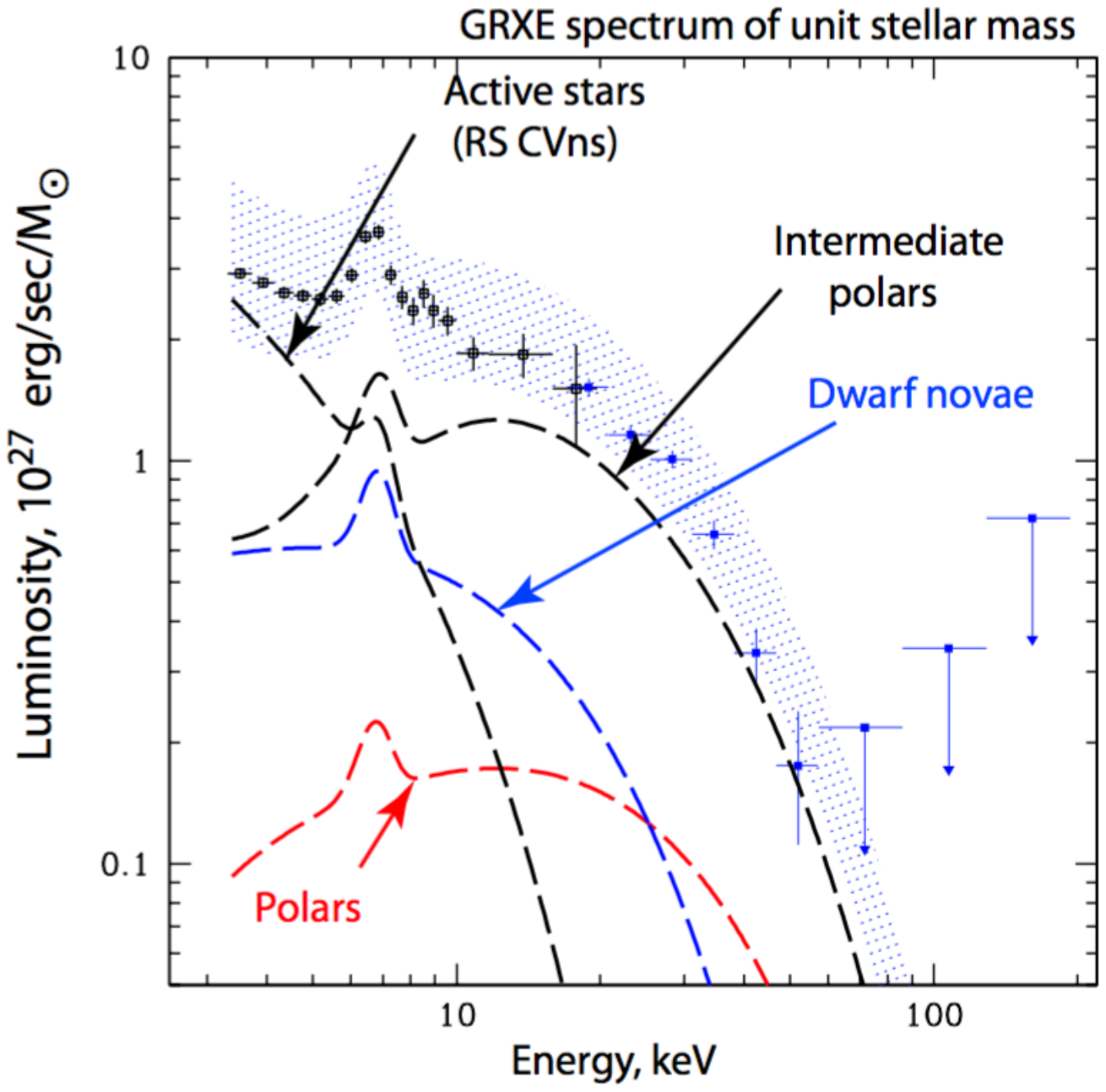}
\caption{GRXE spectrum measured with \emph{RXTE}/PCA (open black squares) and
\emph{INTEGRAL}/IBIS data (filled blue  squares), in comparison with typical
spectra of the X-ray source classes expected to significantly contribute to
the GRXE. These spectra are plotted with normalizations corresponding to
their expected relative contributions to the GRXE (derived from the local
statistics of faint X-ray sources \citep{Sazonov.etal:06}). The shaded region
shows the sum of these spectra with the associated uncertainties. From
\citep{Revnivtsev.etal:07}.
}
\label{fig_grxe_spectra}
\end{figure}

Already early studies based on the IBIS observations showed that the GRXE
spectrum does not extend into the gamma-ray range from the X-ray band with
the same slope \citep{Lebrun.etal:04,Terrier.etal:04}. A later, in-depth
investigation \citep{Krivonos.etal:07} demonstrated that: (i) the GRXE hard
X-ray  (17--60\,keV) surface brightness is proportional to the NIR surface
brightness of the Milky Way (Fig.~\ref{fig_grxe_map}), just like the softer
(3--20\,keV) emission, (ii) the inferred GRXE emissivity per unit stellar
mass in the 17--60\,keV energy band is (0.9--1.2)$\times
10^{27}$\,erg\,s$^{-1} M_\odot^{-1}$, in excellent agreement with the
17--60\,keV luminosity density of CVs in the Solar neighborhood (see the
preceding section), and (iii) there is a pronounced cutoff in the GRXE
spectrum above 20--30\,keV. A comparison of the measured GRXE spectrum with
the sum of the expected contributions of different classes of faint X-ray
sources, weighted according to their relative contributions to the local
X-ray luminosity density, showed an excellent agreement
(Fig.~\ref{fig_grxe_spectra}). The \emph{INTEGRAL} observations have thus
provided strong additional support to the GRXE being the integrated emission
of ABs (dominating at energies below $\sim$5\,keV) and CVs (dominating
between $\sim$5 and $\sim$50\,keV), mainly IPs.

The position of the high-energy cutoff in the GRXE spectrum measured by
\emph{INTEGRAL} can be directly linked to the maximal (virial) temperature of
the X-ray emitting plasma (see Sect.\,\ref{sect:models}) near the surface of
the accreting magnetic WDs composing the GRXE. Based on this idea and the PSR
model spectra from \cite{SRR05}, \citep{Krivonos.etal:07} estimated the
average mass of such objects in the Galaxy to be $\sim$0.5--0.6\,$M_\odot$. A
similar value, 0.6--0.7\,$M_\odot$, was obtained in \cite{Yuasa.etal:12}
based on {\it Suzaku} GRXE observations.

Recently, unresolved extended hard X-ray (20--40\,keV) emission has been
detected by \emph{NuSTAR} in the inner few parsecs of the Galaxy
\citep{Perez.etal:15}. Resolved hard X-ray point sources have been further
identified in the \emph{NuSTAR} survey of 0.6\,deg$^2$ Galactic Center region
\cite{2016ApJ...825..132H}. As demonstrated by \citep{Hailey.etal:16}, the
most natural interpretation of this signal is integrated emission from a
large population of IPs in this region. Interestingly, the inferred typical
mass of the WDs in these systems, $\sim$0.9\,$M_\odot$, is somewhat higher
than that estimated for the whole Galaxy from the \emph{INTEGRAL}
observations of the GRXE (see above), which urges further investigation.
However, it is worth noticing that the high mass value is however not so
different from that of normal CVs (see \cite{2011A&A...536A..42Z}). We note
in this connection that the average WD mass in nearby bright IPs
(0.8--0.9\,$M_\odot$, see \cite{Yuasa.etal:10, SDW19, deMartino.etal:19}) is
close to the value found in \citep{Hailey.etal:16}.

\section{Conclusions}

The contribution of the \int\ observatory in the investigations of
hard X-ray sky is very significant. The surveys of the Galactic plane
and the Galactic center provide a lot of new information about known
X-ray sources and allow to discover many new sources. Accreting white
dwarfs in close (CVs) and wide (SySs) binary systems were also
extensively studied with INTEGRAL. Here we have described the main
results obtained with \int\ in this field.

We presented the identified accreting WDs observed with \int\ until November
2019 although many new identifications will further increase this class of
hard X-ray emitting galactic sources. The most numerous CVs among them are of
the IP type. \int\ discovered 21 new CVs and SySs, including candidates, and
the observed X-ray spectra were used for the determination of the WD masses
in 18 bright IPs and polars. The WD masses in four sources were also
estimated in this review for the first time. We also discussed the
contribution of the \int\ observatory in the study of symbiotic stars.

The \int\ CV sample has permitted the first measurement of the hard X-ray
(17--60\,keV) luminosity function of CVs in the Solar neighborhood between
$10^{32}$ and $10^{34}$\,erg\,s$^{-1}$ (and stringent upper limits on the CV
space density at $10^{31}$--$10^{32}$ and
$10^{34}$--$10^{35}$\,erg\,s$^{-1}$) and of the total hard X-ray luminosity
density of nearby CVs (dominated by IPs): $(1.3 \pm 0.3)\times
10^{27}$\,erg\,s$^{-1}$\,$M_\odot^{-1}$ (17--60\,keV). Thanks to the good
coverage of the Galactic plane and bulge regions by \int\ observations, it
has been demonstrated that the Galactic unresolved hard X-ray emission (from
the Galactic Ridge) closely follows the distribution of stars in the Milky
Way, with the inferred emissivity per unit stellar mass in the 17--60\,keV
energy band of (0.9--1.2)$\times 10^{27}$\,erg\,s$^{-1} M_\odot^{-1}$. The
excellent agreement with the aforementioned hard X-ray luminosity density of
nearby CVs implies that the GRXE is mainly produced by IPs and other CVs at
hard X-ray energies. This conclusion has been further strengthened by the
measurement of the GRXE energy spectrum with \int, which revealed a cutoff
above 20--30\,keV as expected from it being the superposition of CV spectra.

\bigskip
{\bf Acknowledgements.} This review was supported by several programs and
grants. AL, VS and VD acknowledge a support from the Russian Science
Foundation, project 19-02-00423. GJML acknowledges financial support from
grants ANPCYT-PICT 0478/14, PICT 0901/2017. DdM acknowledges financial
support from ASI-INAF contracts I/037/12/0 and .2017-14-H.0 and INAF-PRIN
SKA/CTA Presidential Decree 70/2016 and INAF "Sostegno alla ricerca
scientifica main streams dell'INAF", Presidential Decree 43/2018. LD
acknowledges grant 50 OG 1902. The work of VS was also supported by the DFG
grant WE 1312/51-1.

\bibliography{CV_INT_rev}

\end{document}